\documentclass[useAMS,usenatbib]{mn2e}
\usepackage{amsmath}
\usepackage{amsbsy}
\usepackage{times}
\usepackage{graphicx}
\usepackage{aas_symbols}
\usepackage{threeparttable}
\usepackage{epstopdf}
\usepackage{url}
\usepackage{cases}
\usepackage{color}
\usepackage{multirow}

\begin{document}
	
	\title[Successful short GRB jets]
	{A lesson from GW170817: most neutron star mergers result in tightly collimated successful GRB jets}
	
	\author[Beniamini et al.]{Paz Beniamini$^{1,2}$\thanks{Email: paz.beniamini@gmail.com}, Maria Petropoulou$^3$, Rodolfo Barniol Duran$^4$ $\&$ Dimitrios Giannios$^{5}$  \\
		$^{1}$Department of Physics, The George Washington University, Washington, DC 20052, USA \\
		$^2$Astronomy, Physics and Statistics Institute of Sciences (APSIS)\\	
		$^3$Department of Astrophysical Sciences, Princeton University, 4 Ivy Lane, Princeton, NJ 08544, USA\\
		$^4$Department of Physics and Astronomy, California State University, Sacramento, 6000 J Street, Sacramento, CA 95819, USA\\
		$^{5}$Department of Physics and Astronomy, Purdue University, 525 Northwestern Avenue, West Lafayette, IN 47907, USA
	}
	
	\date{Accepted; Received; in original form ...}
	
	\pubyear{2018}
	
	\maketitle
	
	\begin{abstract}
		The joint detection of gravitational waves (GWs) and $\gamma$-rays from a binary neutron-star (NS) merger provided a unique view of off-axis GRBs and an independent measurement of the NS merger rate. Comparing the observations of GRB170817 with those of the regular population of short GRBs (sGRBs), we show that an order unity fraction of NS mergers result in sGRB jets that breakout of the surrounding ejecta. We argue that the luminosity function of sGRBs, peaking at $\approx 2\times 10^{52}\, \mbox{erg s}^{-1}$, is likely an intrinsic property of the sGRB central engine and that sGRB jets are 
		typically narrow with opening angles $\theta_0 \approx 0.1$. We perform Monte Carlo simulations to examine models for the structure and efficiency of the prompt emission in off-axis sGRBs. We find that only a small fraction ($\sim 0.01-0.1$) of NS mergers detectable by LIGO/VIRGO in GWs is expected to be also detected in prompt $\gamma$-rays and that GW170817-like events are very rare. For a NS merger rate of $\sim 1500$ Gpc$^{-3}$ yr$^{-1}$, as inferred from GW170817, we expect within the next decade up to $\sim 12$ joint detections with off-axis GRBs for structured jet models and just $\sim 1$ for quasi-spherical cocoon models where $\gamma$-rays are the result of shock breakout. Given several joint detections and the rates of their discoveries, the different structure models can be distinguished. In addition the existence of a cocoon with a reservoir of thermal energy may be observed directly in the UV, given a sufficiently rapid localisation of the GW source.
	\end{abstract}
	\section{Introduction}
	The connection between neutron star-neutron star (NS-NS) or neutron star-black hole (NS-BH) mergers with short-duration Gamma Ray Bursts (sGRBs) and the nucleosynthesis of $r$-process elements dates back to a few seminal works from the nineteen seventies and eighties \citep{Lattimer1974,Lattimer1976,Blinnikov1984,Paczynski1986,Goodman1986,Eichler1989}. In recent years, the rate of $r$-process formation has been constrained using various observational lines of argument \citep{Hotokezaka2015,ji2016Nature,Beniamini2016a,Macias2016,HBP2018,BDS2018}. It was shown to be broadly consistent with the rate of sGRBs \citep{Guetta2006,Guetta2009,Coward2012,wanderman_piran2015,Ghirlanda2016} and with the rate of NS mergers as inferred from observations of Galactic double neutron stars \citep{Kochanek1993,Kim2015}. However, a clear determination of NS-NS mergers being the progenitors of sGRBs and the main source of $r$-process elements remained somewhat uncertain until the recent discovery of the kilonova AT2017gfo \citep{Tanvir2017} and GRB170817 \citep{Goldstein2017}, accompanying the NS-NS merger event GW170817 \citep{GW170817}. This recent discovery also leads to a new and independent estimate of the rate of NS-NS mergers. Furthermore, the observations of a very weak GRB accompanying the event, and detailed modelling of the peculiar and long-lived afterglow that followed the event, provides us with information regarding the opening angle, viewing angle, and core luminosity of GRB170817.

	The GW170817/GRB170817 event provides an unprecedented opportunity to explore sGRB jets in a broader context:
	How frequently do sGRB jets manage to break through the ejecta material surrounding the NS-NS mergers? What are the typical opening angles of sGRB jets? What determines the shape of the sGRB luminosity function? These topics have been partially explored in the past, either from an observational point of view, using the data from electromagnetically-detected sGRBs (i.e., not accompanied by a GW event) \citep{wanderman_piran2015,Ghirlanda2016,Moharana2017}, or from hydrodynamical studies of the GRB jet propagation through the NS-NS merger ejecta \citep{Aloy2005,Hotokezaka2013,Nagakura2014,Murguia-Berthier2014,Lazzati2017,Bromberg2018,Duffell2018}. Here, we show that by combining this knowledge with the unique constraint from GW170817 we can probe these questions in greater detail, thus significantly improving our understanding of these issues.

	The discovery of GRB170817 raised perhaps as many new questions as answers. In particular, the extremely dim prompt GRB accompanying the event has led to different interpretations, according to which the observed $\gamma$-ray emission is either arising from the `wings' of the jet (material beyond the core that is less energetic) due to the same physical process that produces the prompt emission along the jet's axis \citep{Kathirgamaraju2018,Lamb2017,GGG2017}, or is due to shock breakout from the thermal energy stored in the cocoon that was produced by the jet-ejecta interaction \citep{Lazzati2017,Gottlieb2017,Kasliwal2017}. Although GRB170817 alone is not enough to distinguish between the proposed models, we show that our understanding can be significantly improved once a sample of joint GRB and GW detections has been established and their luminosity and viewing angle distributions have been studied.

	The paper is organized as follows. In \S \ref{sec:failfrac} we revisit the rate and luminosity function of sGRBs and show that the latest results from GW170817 strongly constrain: the fraction of  NS-NS mergers accompanied by sGRBs, the opening angles of sGRBs, and the interpretation of their luminosity function. In \S \ref{sec:GRBwjointdet} we consider different models for the off-axis prompt GRB emission. By performing Monte Carlo simulations, we compare the observed properties arising from different emission models, in cases with a joint GW and prompt GRB detection. In \S \ref{sec:cocooncooling} we focus on the cocoon model and discuss an additional observable signal that can directly constrain the thermal energy stored in the cocoon. Finally, we discuss some implications of this work and conclude in \S \ref{sec:discussion}.
	
	\section{Successful sGRB jets and the sGRB luminosity function}
	\label{sec:failfrac}
	\subsection{General considerations - rate comparison}
	The observed luminosities of sGRBs are a convolution of the intrinsic luminosity function $\Phi(L)$ 
	and the volumetric rate $R(z)$. 
	\cite{wanderman_piran2015} find that, in the local universe, the rate of sGRBs is:  
	\begin{equation}
	\label{eq:WPrate}
	R_{\rm sGRB}= 4.1^{+2.3}_{-1.9}f_b^{-1}L_{\rm m,49.7}^{-\alpha_L} \, \, \mbox{Gpc}^{-3} \mbox{ yr}^{-1}
	\end{equation}
	where $\alpha_L = 0.95$, $L_{\rm m,49.7}\equiv L_{\rm min}/(5\times 10^{49} \mbox{erg s}^{-1})$, $f_b$ is the beaming factor, and $L_{\rm min}$ is the minimum luminosity of the GRB luminosity function, which is typically assumed. Note that the lowest observed luminosity of a sGRB in the sample considered by \cite{wanderman_piran2015} is $\approx 2\times 10^{50}\mbox{ erg s}^{-1}$. This is an absolute maximum on the true value of $L_{\rm min}$.

	It has been long theorized \citep{Blinnikov1984,Paczynski1986,Goodman1986,Eichler1989} and recently proved by the  dual detection of GW170817/GRB170817 \citep{GW170817} that NS mergers (either NS-NS or NS-BH) lead to sGRBs. However, it is possible that not all NS mergers result in a successful GRB.  We denote the ratio of failed to successful sGRBs rates by $r_{\rm fail}\equiv R_{\rm fail}/ R_{\rm sGRB}$. The NS merger rate as inferred by the LIGO event, GW170817, is $R_{\rm merg} = 1540^{+3200}_{-1220}$ Gpc$^{-3}$ yr$^{-1}$ \citep{GW170817}. Combining this with equation~(\ref{eq:WPrate}) and assuming a double-sided top-hat jet with opening angle $\theta_0\approx 0.1 \theta_{0, .1}$, we find: 
	\begin{equation}
	r_{\rm fail}=
	\frac{R_{\rm merg}}{R_{\rm sGRB}}-1\approx 2\, L_{\rm m,49.7}^{\alpha_L} \theta_{0,.1}^2R_{\rm m,1540}-1
	\label{eq:rates}
	\end{equation}
	where $R_{\rm m,1540}\equiv R_{\rm merg}/(1540 \mbox{ Gpc}^{-3}\mbox{ yr}^{-1})$ and the numerical value holds for $\alpha_L=0.95$. Note that for clarity we have dropped here and in the following the errors in $R_{\rm merg}$, but our main conclusions remain unaltered, when considering the minimum and maximum allowed value of $R_{\rm merg}$.
	Since neither $L_{\rm min}$ can be larger than $5\times 10^{49}\mbox{erg s}^{-1}$ nor $\theta_0$ can change significantly compared to the canonical value used above, the ratio of failed to successful jets cannot be very large. On the contrary, this ratio is estimated to be $r_{\rm fail}\sim 4\times 10^2$ in long (collapsar) GRBs; here we apply equation \ref{eq:rfail} but with $\alpha_L=0.2,\beta_L=1.4, L_{\rm min}=2\times 10^{50}\mbox{erg s}^{-1}, L_*=3\times 10^{52}\mbox{erg s}^{-1}$, as appropriate for long GRBs \citep{WP2010} \citep[see also][]{mariaetal2017,Sobacchi2017}. This huge difference in the expected rates of failed short and long GRBs may be related to the nature of the ejecta that the jet has to propagate through. For example, the expansion of the dynamical ejecta  in the case of a NS merger may facilitate the jet breakout \citep{Duffell2018} in contrast to the jet propagation through the quasi-static outer layers of the collapsing star expected in the case of long GRBs \citep{Bromberg2011}.

	\subsection{Interpretation of the sGRB luminosity function}
	\label{sec:lumfun}
	The (isotropic equivalent) luminosity function of both short and long GRBs (after de-convolving with the rate function) is described by a broken power law:
	\begin{equation}
	\label{eq:pgrb_intrinsic}
	\phi(L) \equiv \frac{{\rm d}N_{\rm GRB}}{{\rm d}\log L} = \left \{ 
	\begin{array}{ll}
	\left(\frac{L}{L_*}\right)^{-\alpha_L} & L_{\rm min}\le L \le L_*  \\
	\left(\frac{L}{L_*}\right)^{-\beta_L} & L > L_*.
	\end{array}
	\right.
	\end{equation}
	For sGRBs $\alpha_L \approx 0.95$, $\beta_L \approx 2.0$ and $L_* \approx 2 \times 10^{52} \mbox{erg s}^{-1}$\citep{wanderman_piran2015}. Henceforth, we adopt the luminosity function reported by \cite{wanderman_piran2015}, while noting that none of our main conclusions would change, if we were to adopt other values for the power-law slopes and/or break luminosity reported in the literature \citep{Guetta2006,Salvaterra2008,Ghirlanda2016}. We next explore three different interpretations to the broken power-law nature of $\Phi(L)$: (i) arising due to the increasing fraction of failed GRB jets for $L\le L_*$, (ii) being the result of the angular structure of sGRBs, or finally (iii), being an intrinsic property of the sGRB central engine.  
	
	\subsubsection{Failed jets}
	\label{sec:failedjets}
	\cite{mariaetal2017} have argued that the broken power-law nature of $\Phi(L)$ in long GRBs could result from an underlying single power-law distribution of luminosities, that is then modified at lower luminosities due to an ever increasing fraction of failed GRB jets. This provides a natural explanation for the apparent more complicated luminosity function whereby the interpretation for $L_*$ is the maximum luminosity for which not all long GRB jets manage to break out of the stellar envelope.

	Interestingly, the same interpretation cannot hold for sGRBs. This is clear from the fact that within the \cite{mariaetal2017} framework, the slope of $\Phi(L)$ above $L_*$ is that of the true underlying distribution. In this context, the  ratio of failed sGRB to successful sGRB engines at $L < L_{*}$ can be then extrapolated to lower luminosities as: 
	\begin{equation}
	\label{eq:rfail}
	r_{\rm fail}=\frac{R_{\rm merg}}{R_{\rm sGRB}} -1 \approx \left(\frac{L_{\rm min}}{L_{*}}\right)^{-\beta_L+\alpha_L}\approx 540 \,  L_{\rm m,49.7}^{-1} ,
	\end{equation}
	which is clearly in contradiction with equation (\ref{eq:rates}) by more than two orders of magnitude. 
	
	\subsubsection{Angular structure}
	\label{sec:angular}
	Another potential way to explain the broken power-law nature of the luminosity function is to consider that GRBs have a wide structure beyond the jet core, i.e., the jet luminosity is constant up to some $\theta_0$ and then decreases for larger angles \citep{Lipunov2001,Frail2001,Rossi2002,Zhang2002,GK2003,Eichler2004,VanEerten2012,Pescalli2015}. In this context, the distribution $\Phi(L)$ below $L_*$ may become dominated by GRBs that are marginally above the critical core luminosity $L_*$, but are seen progressively  more off axis as $L\ll L_*$. This particular scenario would imply an increased intrinsic minimal core luminosity, $L_{\rm min}\approx L_*$, and hence would increase the ratio of failed to successful jets given by equation (\ref{eq:rates}), which could help to reduce (and potentially resolve) the inconsistency with the estimate given by equation (\ref{eq:rfail}).  As an illustration, let us consider that the jet (isotropic equivalent) luminosity is constant up to a latitude $\theta_0$ and then declines as $\theta^{-\delta}$ for $\theta>\theta_0$. Assuming an isotropic distribution of solid angles and the same structure for all sGRB jets, this will lead to $\Phi(L)\propto L ({\rm d}N/{\rm d}\theta) ({\rm d}\theta/{\rm d}L) \propto L^{-2/\delta}$. This matches the observed $\Phi(L)\propto L^{-\alpha_L}$, only if the typical value of the jet structure slope is $\delta \approx 2$. Such a shallow profile however (close to the so called ``universal jet'' structure; see \citealt{Rossi2002,Zhang2002}) could be in contention with observations of the X-ray afterglow luminosity to $\gamma$-ray energy ratio, unless the Lorentz factors of GRBs remain $\gtrsim 50$ at large angles \citep{BN2018}.

	It is constructive to compare the value of $\delta$ obtained above, with the constraints from GRB170817. Unless the efficiency of $\gamma$-ray production decreases significantly at angles beyond the core (in which case the angular structure hypothesis for the luminosity function is clearly invalid), the observed $\gamma$-ray luminosity of that event,  $L_{\rm GRB170817}$, must be at least as large as the extrapolation of the core luminosity $L_0$ to the observed angle (so as not to over-produce $\gamma$-rays compared to observations). This leads to a lower limit on $\delta$, i.e.  $\delta \geq \log(L_{\rm GRB170817}/L_0)/\log(\theta_0/\theta_{\rm obs})$, where $L_{\rm GRB170817}=1.6\times 10^{47} \mbox{erg s}^{-1}$ \citep{Goldstein2017} and $\theta_{\rm obs}\approx 0.35$ \citep{Mooley2018B}.
	Taking $L_0\approx L_*$, which is roughly the minimal core luminosity under the interpretation that the shape of the luminosity function is driven by the angular structure, and adopting the minimum allowed value for $\theta_0\approx 0.05$ (see \S \ref{sec:angular}), we find $\delta \gtrsim 6$; larger values of $\theta_0$ would require even steeper profiles. The obtained value of $\delta$ is in contention with $\delta\approx 2$ derived previously.  
	
	Thus, the interpretation that the luminosity below $L_*$ is dominated by off-axis events is challenged by observations, under the assumptions of a common jet structure among all sGRBs and large efficiency beyond the jet core.
	
	\subsubsection{Intrinsic structure}
	A third possibility is that the broken power-law luminosity function determined by observational studies \citep{Guetta2006,Guetta2009,wanderman_piran2015,Ghirlanda2016} actually reflects the intrinsic power produced by the central engine. We turn next to explore the implications of this interpretation of the luminosity function on the typical opening angles of sGRBs and on the expected distributions of luminosities and observation angles as probed by future joint detections of prompt GRBs and GWs.

	\subsection{Constraining the opening angles of short GRBs}
	\label{sec:angles}
	We now use the connection between the rate of NS mergers and sGRBs to constrain the jet opening angle, $\theta_0$, as described below.
	Following our reasoning in \S \ref{sec:angular}, we assume here that there is a typical (or universal) sGRB opening angle, and that sGRB angular profiles are sufficiently steep, such that they are typically observed at $\theta_{\rm obs}\lesssim \theta_0$.
	Different arguments and constraints point to jets having a narrow range of opening angles centered around $\theta_0 \approx 0.1$.

	First, since $R_{\rm merg}\ge R_{\rm sGRB}$, equation~(\ref{eq:rates}) provides a {\it lower} limit on the allowed opening angles of sGRBs (or equivalently on the beaming factor) which is roughly: 
	\begin{equation}
	\theta_0\ge \theta_{\rm min}\equiv 0.07 \, L_{\rm m,49.7}^{-\alpha_L/2}R_{\rm m, 1540}^{-1/2}
	\label{eq:thmin}
	\end{equation}

	At the same time,  observations of sGRBs over the last 14 years with {\it Neil Gehrels} {\it Swift} satellite, provide us with an {\it upper} limit on the allowed value for sGRB opening angles. Considering the sample of observed {\it Swift} BAT bursts with $T_{90}<2$~s \footnote{This choice does not result in a pure non-collapsar GRB sample. The
		contamination by collapsar GRBs, however, can only increase $P_{\max}$ and in turn $\theta_{\max}$ with respect to its true value.}, we search the online database\footnote{\url{https://swift.gsfc.nasa.gov/archive/grb_table/}} and find that the maximum peak photon flux of sGRBs within $\Delta T_{\rm obs}=14$ years of observations (in the $15-150$~keV range) is $P_{\rm max}=12.1 \mbox{ ph cm}^{-2}\mbox{ s}^{-1}$. For a given redshift $z$ and corresponding luminosity distance $d_L(z)$, this can be translated to a $\gamma$-ray peak luminosity as: 
	\begin{equation}
	L(z,P_{\rm max})=\frac{4\pi d^2_L(z)}{1+z}\frac{\int_{1\rm keV}^{10 \rm MeV}EN(E){\rm d}E}{\int_{15(1+z)\rm keV}^{150(1+z) \rm keV}N(E){\rm d}E}P_{\rm max},
	\label{eq:LfromPmax}
	\end{equation}
	where $N(E)$ is the differential photon spectrum described by the so-called Band function (computed in the source frame) \citep{Band1993}. Here, we have considered the spectral range at the source to be 1~keV-10~MeV. The latter is the same spectral range over which the luminosities in \cite{wanderman_piran2015} were computed and is therefore the appropriate choice in order to make a comparison with their values, as we do below. In what follows we assume typical observed sGRB parameters: $\alpha=-0.5$, $\beta=-2.25$, $E_{\rm p, source}=500$~keV \citep{Nava2011}. Using equation (\ref{eq:LfromPmax}) and requiring that $L(z,P_{\rm max})>L_{\rm min}=5\times 10^{49}\mbox{ erg s}^{-1}$ (the same value of $L_{\rm min}$ considered above), we find a conservative minimum distance from which current {\it Swift} BAT bursts could have originated $d_L(z)\equiv d_{P_{\rm max}}\ge 250$~Mpc. In other words, even a burst that had an intrinsic (isotropic equivalent) luminosity at the core as low as $L_{\rm min}\approx 5\times 10^{49}\mbox{ erg s}^{-1}$ would have resulted in a peak flux larger than {\it any} of the observed {\it Swift} BAT bursts, if its distance from us was less than 250 Mpc. Thus, the opening angle of sGRBs has to be small enough to be consistent with the non-detection of bursts with $L>L_{\rm min}$ at $d_L(z)< 250$~Mpc.  For a fixed rate of NS mergers, the upper limit on $\theta_0$ is independent of the inferred local sGRB rate given by equation (\ref{eq:WPrate}):
	\begin{eqnarray}
	\label{eq:thmax}
	& \theta_{\rm max}\!=\sqrt{\frac{2(r_{\rm fail}+1)}{R_{\rm merg}\Delta T_{\rm obs}V_{\rm max}\eta_{\rm Sw}}}= 0.1\,(r_{\rm fail}+1)^{\frac{1}{2}} \times  \\
	& R_{\rm m, 1540}^{-1/2} \bigg(\frac{14 \mbox{ yr}}{\Delta T_{\rm obs}}\bigg)^{\frac{1}{2}} \bigg(\frac{250 \mbox{ Mpc}}{d_{P_{\rm max}}}\bigg)^{\frac{3}{2}} \bigg(\frac{0.12}{\eta_{\rm Sw}}\bigg)^{\frac{1}{2}}, \nonumber
	\end{eqnarray}
	where $V_{\rm max}\equiv (4\pi/3) \, d_{P_{\rm max}}^3$ and where we have conservatively used an efficiency of $\eta_{\rm Sw}=0.12$ for the detectability of {\it Swift} sGRBs above the BAT threshold \citep{Burns2016}. We note also that a comparable, though somewhat larger estimate for $\theta_{\rm max}$ can be obtained by using {\it Fermi} GBM data. In this case, the efficiency is larger $\eta_{\rm GBM}=0.6$ \citep{Racusin2017} while the observation period is slightly shorter $\Delta T_{\rm obs}=9$ years. According to \cite{Lu2017}, $P_{\rm max}=138\,\mbox{ph cm}^{-2}\mbox{ s}^{-1}$ (between 8~keV and 40~MeV) for this period. Using the typical spectral parameters for GBM bursts, namely $\alpha=-1.5, E_p=600$~keV \citep{Nava2011}, we find $d_{P_{\max}}=130$~Mpc and $\theta_{\rm max}^{(\rm GBM)}=0.16$.

	As we show next, comparing the ratio of merger rates to sGRB rates as given by equation (\ref{eq:rates}) to the fraction of successful jets as probed by recent numerical analysis of the jet-merger outflow interaction \citep{Duffell2018} provides in turn, an {\it estimate}  of $\theta_0$ (instead of a lower and upper limit as presented above).
	
	\cite{Duffell2018} find that jets are successful if the beaming corrected energy in the jet $E_{\rm j}$ is larger than a critical value $E_{\rm cr}\equiv 0.05 \theta_0^2 E_{\rm ej}$, where $E_{\rm ej}$ is the energy deposited in the NS merger ejecta. Throughout this work, we assume that the opening angle of the injected jet in the simulations by \cite{Duffell2018} is equal to the jet opening angle after breakout from the NS merger ejecta.  This is an appropriate approximation for jets propagating in a non-expanding medium \citep{Bromberg2011}, but it remains to be shown whether this also holds for jets pummeling through an expanding medium. Taking $E_{\rm ej}=10^{51}$~erg as a typical value for the kinetic energy of the NS merger ejecta \citep{HBP2018}\footnote{This is inferred from the modelling of the kilonova emission following GW170817 \citep{Kasen2017,Tanaka2017}.} and a constant time of engine activity given by the typical sGRB duration (in the central engine frame) $t_{\rm e}=0.3 \ t_{\rm e, .3}$~s \citep{Kouveliotou1993}, the condition $E_{\rm j}> E_{\rm cr}$ translates to:
	\begin{equation}
	L>L_{\rm cr}\simeq 5\times 10^{49}\frac{E_{\rm ej,51}}{t_{\rm e, .3}} \bigg(\frac{\eta_{\gamma}}{0.15}\bigg)\mbox{erg s}^{-1},
	\label{Lcr}
	\end{equation}
	where $E_{\rm ej,51}\equiv E_{\rm ej}/(10^{51}\, \mbox{erg})$, $\eta_{\gamma}$ is the efficiency of converting the injected jet power into observed $\gamma$-rays in the prompt phase and is normalized to a characteristic value $\eta_{\gamma}=0.15$ inferred from observations \citep{Beniamini2015,Beniamini2016}. Notice that $L_{\rm cr}$ is very close to the canonical value for $L_{\rm min}$. Were $L_{\rm cr}\gg L_{\rm min}$, then a second break in the sGRB luminosity function should be present, which is not observed\footnote{The possibility of $L_{\rm cr}=L_*$ is ruled out by the rate comparison in equation \ref{eq:rates}, as was demonstrated explicitly in \S \ref{sec:failedjets}.}. The ratio of failed to successful sGRB rates can also be estimated from the sGRB luminosity function (see equation (\ref{eq:pgrb_intrinsic})) and it may be written as $r_{\rm fail}\approx (L_{\rm cr}/L_{\rm min})^{\alpha_L}-1\approx 1$. This condition is satisfied when
	\begin{equation}
	\theta_0=0.07 \bigg(\frac{E_{\rm ej,51}}{t_{\rm e, .3}}\frac{\eta_{\gamma}}{0.15}\bigg)^{\frac{\alpha_L}{2}}  L_{\rm m,49.7}^{-\alpha_L} R_{\rm m, 1540}^{-1/2},
	\label{eq:theta}
	\end{equation}
	where equation (\ref{eq:rates}) was used. Interestingly, the allowed values of $\theta_0$ obtained from equations (\ref{eq:thmin}), (\ref{eq:thmax}), and (\ref{eq:theta}) are consistent with two independent estimates of the opening angle in GRB170817:
	\begin{enumerate}
		\item recent measurements of superluminal motion in GRB170817 \citep{Mooley2018B} using VLBI data demonstrated that $\theta_0\lesssim 0.09$ in GRB170817.
		\item the peak time of the GRB170817 afterglow in X-rays and radio suggested that $0.04\lesssim \theta_0 \lesssim 0.1$ \citep{Mooley2018,Pooley2018}.
	\end{enumerate}
	Our analysis suggests that the majority of sGRB jets (or even all) is successful in breaking out of the surrounding ejecta. In addition, assuming that the opening angle of sGRBs is independent of their luminosity, we showed that $\theta_0 \approx 0.1$; these are tighter constraints than existing ones for the sGRB population \citep{Janka2006,Fong2015}. Note that these results do not depend too strongly on the value of $R_{\rm merg}$, which will be better constrained in the future with more observations of GWs from NS mergers \citep[see also][for comparison with $r$-process rates suggesting a comparable but possibly slightly lower rate]{Hotokezaka2018}.

	\section{Probing GRB models with joint GW \& prompt GRB detections}
	\label{sec:GRBwjointdet}
	The detection of a GW signal could boost the likelihood of association of an otherwise marginal $\gamma$-ray signal with an underlying GRB and of the identification of the host galaxy \citep{Patricelli2016,joint170817, Kathirgamaraju2018,Beniamini2018}. NS mergers detected by GWs also select nearby events. The observed distribution of sGRB luminosities that are also detected in GWs is therefore expected to be different than that of sGRBs with no GW detection. This implies that joint GW and  prompt GRB detections could provide a unique probe to the distribution of the $\gamma$-ray luminosity as a function of angle $\theta$ with respect to the jet symmetry axis $L_{\gamma,\rm iso}(\theta)$, and therefore, to the underlying jet and/or cocoon physics.

	Indeed, the observation of the first sGRB with a GW signal is a demonstration of this point. This GRB had a prompt $\gamma$-ray luminosity which is approximately four orders of magnitude lower than the previously weakest sGRB with confirmed redshift and it was observed from a relatively large viewing angle, i.e., $\theta_{\rm obs}>\theta_0$ \citep{Finstad2018}.  In this case, the low observed luminosity may be the result of a quasi-isotropic radiatively inefficient\footnote{This is compared to the available energy along the line of sight to the observer.} component (e.g., due to a cocoon) that accompanies GRB jets or of the intrinsic angular jet structure, which could suppress the emission at angles $\theta \gtrsim \theta_0$ \citep{Kathirgamaraju2018,Fraija2017,Lamb2017,Lazzati2017,Lazzati2017B,Gottlieb2017,Kasliwal2017,Margutti2018,Troja2018,Barkov2018,Lazzati2018}.

	\begin{figure}
		\centering
		\includegraphics[width = .47\textwidth]{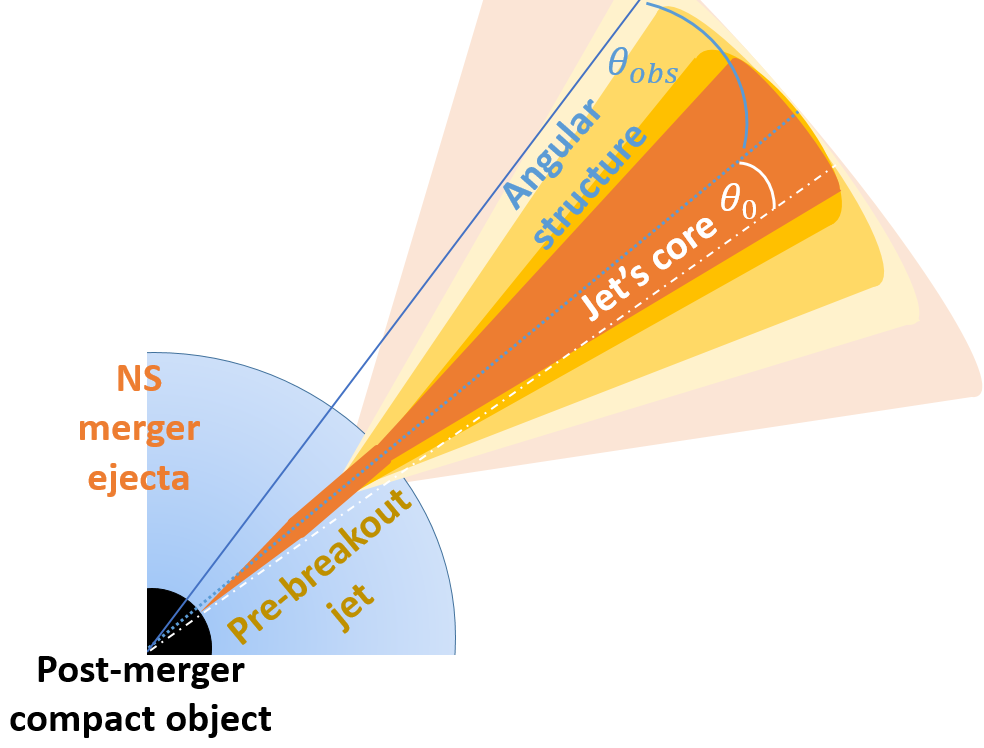}
		\includegraphics[width = .47\textwidth]{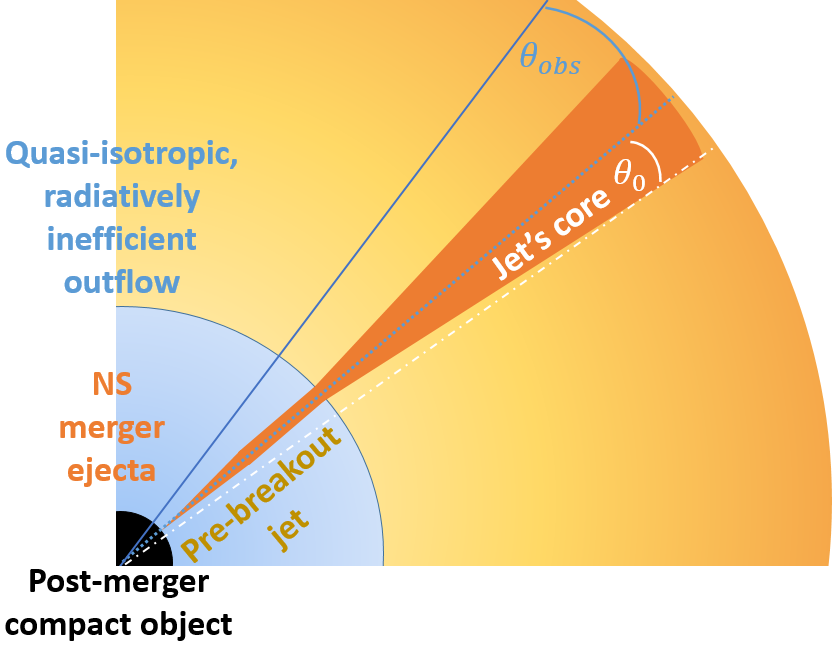}
		\caption{A schematic illustration of a short GRB jet that was successful in breaking out of the NS merger ejecta. Top: A structured jet with a strong angular profile beyond the jet core defined by angle $\theta_0$. An off-axis observer (located at an angle $\theta_{\rm obs}$) receives emission from the less energetic material (compared to the jet core) moving along their line of sight (SJ model). Bottom: The jet core is surrounded by a radiatively inefficient quasi-isotropic component with a likely radial structure in its properties (CL model). } 
		\label{fig:illustrate}
	\end{figure}  
	\subsection{Monte Carlo simulations}\label{sec:MC}
	We focus here on two extreme cases for the angular/radial profile of the jet: (i) a `structured- jet' (SJ) model, in which the luminosity is a power-law (PL) or Gaussian (GS) function of the angle from the core and the radiative efficiency remains constant as a function of latitude and (ii) a `cocoon-like' (CL) model, where a top-hat jet is accompanied by a quasi-spherical, radiatively inefficient outflow, possibly with significant radial structure. A schematic illustration is presented in Figure \ref{fig:illustrate}. We stress that in reality the situation may lie somewhere in between these two extremes.

	We perform a Monte Carlo (MC) simulation to explore the resulting $\gamma$-ray luminosity, $L_{\gamma,\rm iso}$ and observation angle, $\theta_{\rm obs}$, distributions arising from different parameters of the two underlying models mentioned above. 
	We begin each realization of the MC  by drawing a random orientation between the jet  core (which is taken to be perpendicular to the merger plane) and the line of sight (i.e. uniform in $\cos \theta$ and $\phi$) and a distance, $d$, of the event according to the rate function $R(z)$ of \cite{wanderman_piran2015}. We impose a cut-off at the maximum distance for which NS mergers can be detected in GWs with future capabilities, $d_{\rm max}=220$~Mpc \citep{GW170817}\footnote{For the distance scales relevant for GW detection, cosmological effects can, to a good approximation, be neglected, and the bursts can be assumed to be randomly distributed within a sphere of radius $d_{\rm max}$.}; later on we test the dependence of the results on $d_{\rm max}$, by adopting $d_{\rm max}=50$~Mpc for all explored models. Following our reasoning in \S \ref{sec:lumfun}, we then simulate the luminosity of the jet core according to the luminosity function in (\ref{eq:pgrb_intrinsic}). From the discussion in \S \ref{sec:angles}, we expect $\theta_0\approx 0.1$. We consider this as the canonical value for all simulated bursts, but for completeness, consider also the cases of $\theta_0=0.05$ and $\theta_0=0.2$. For a given structure model, the emissivity is then extrapolated to larger angles. For each event, we calculate both the on line-of-sight contribution to the emission (i.e., from emitters that see the observer within their relativistic beaming cone) and the off line-of-sight contribution \citep{Eichler2004,GranotRamirez2010,Kasliwal2017,Granot2017,Ioka2018,Kathirgamaraju2018}. The latter is calculated in the following way (see, e.g., \citealp{Kasliwal2017}). Assuming $\Gamma\gg \theta_0^{-1}$, the observed energy scales as:
	\begin{eqnarray}
	\frac{E_{\gamma,\rm obs}(\theta_{\rm obs})}{E_{\gamma,\rm em}(\theta_0)}= 
	\left\{ \begin{array}{ll}1 & \! \! \theta_{\rm obs}< \theta_0\\
	\max[\frac{ E_{\gamma,\rm em}(\theta_{\rm obs})}{ E_{\gamma,\rm em(\theta_0)}},q^{-4}] \! \! & \! \! \theta_0<\theta_{\rm obs}\!<\! 2\theta_0\\ \max[\frac{ E_{\gamma,\rm em}(\theta_{\rm obs})}{ E_{\gamma,\rm em}(\theta_0)},q^{-6}(\theta_0\Gamma)^2]\!\! &\! \! 2\theta_0\!<\!\theta_{\rm obs},
	\end{array} \right.  \nonumber
	\end{eqnarray}
	where the sub-script `em' denotes emitted quantities, the sub-script `obs' denotes observed ones, $q=(\theta_{\rm obs}-\theta_0)\Gamma$ and  we set $\Gamma=100$ in our simulations. The duration scales as $\Delta t_{\rm obs}=q^2\Delta t_{\rm em}$ if the burst duration is set by a `single' emission episode that takes place at some location $r$ that extends over a range of radii $\Delta r\lesssim r$ or as $\Delta t_{\rm obs}=\Delta t_{\rm em}$ if the burst duration is set by the radial width of the outflow. In what follows, we assume that the duration is set by the radial width of the outflow and that $\Delta t_{\rm obs}=\Delta t_{\rm em}=0.3$~s. The luminosity then scales as the ratio of $E_{\gamma,\rm obs}$ to $\Delta t_{\rm em}$. This provides an upper limit on the off line-of sight luminosity and is therefore a conservative choice for our simulations. Even in this case, we find that the off line-of-sight emission is sub-dominant to the on line-of sight emission, unless the angular profile is extremely steep and the observation angle is close to jet core.

	Computing the emission from structured-jet (SJ) simulated events is meaningful only if the jet has successfully broken out of the ejecta.This is in accord with results of hydrodynamical simulations of jet and NS merger interaction \citep{Duffell2018}. Furthermore, as shown in \S \ref{sec:failfrac}, a large fraction of sGRB jets (and possibly all) are expected to successfully break out of the expanding ejecta. We therefore assume $r_{\rm fail}=1$ in all SJ models. We explore three variants of the SJ models described below:
	\begin{itemize}
		\item A power-law (SJPL) model, where $L=L_0$ for $\theta<\theta_0$ and $L=L_0(\theta/\theta_0)^{-\delta}$ for $\theta>\theta_0$. The value of $\delta$ in this model is constrained by the observations of GRB170817. For a given value of $\theta_0$, $\delta$ must satisfy\footnote{This condition is at first glance similar to the condition discussed in \S \ref{sec:angular} when considering whether it is possible to explain the broken power-law nature of the luminosity function with angular structure (in which case the minimum core luminosity is given by $L_*$). However, since we argue that this possibility is unlikely and that the luminosity function $\Phi(L)$ is intrinsic, we conservatively replace here $L_*$ by $L_{\rm min}$.} $\delta \geq \log(L_{\rm GRB170817}/L_{\rm min})/\log(\theta_0/\theta_{\rm obs})$, since the lowest core luminosity is given by $L_{\rm min}$ and the extrapolation of this luminosity to $\theta_{\rm obs}\approx 0.35\pm 0.09$ \citep{Mooley2018B,vanEerten2018} should not overproduce $\gamma$-rays as compared with those observed for GRB170817. This consideration implies $\delta \geq 3$ for $\theta_0=0.05$, $\delta \geq 4.5$ for $\theta_0=0.1$ and $\delta \geq 10$ for $\theta_0=0.2$ (where we have taken here into account the uncertainty $\theta_{\rm obs}$ quoted above). 
		\item A Gaussian (SJG) model, where $L=L_0 \exp[-(\theta/\theta_0)^2]$. As in the SJPL case, this family of models can be limited by the requirement to not overproduce $\gamma$-rays compared to GRB170817. Applied to this model, this consideration puts an upper limit on $\theta_0$ of: $\theta_0\leq \theta_{\rm obs}(\log(L_{\rm min}/L_{\rm GRB170817}))^{-1/2}\approx 0.14$  (where once more we have taken here into account the uncertainty $\theta_{\rm obs}$ quoted previously).
		\item A simplified cocoon-like (CL) model, where $L=L_0+L_{\gamma,\rm co}$ for $\theta_{\rm obs}<\theta_0$ and $L=L_{\gamma,\rm co}$ for $\theta_{\rm obs}>\theta_0$. Here, $L_{\gamma,\rm co}$ is the peak (isotropic equivalent) cocoon breakout luminosity. An estimate of this luminosity (and related observational consequences of the thermal energy reservoir stored in the cocoon) is provided in \S \ref{sec:cocooncooling} (and in particular equation \ref{Lgco}).

	\end{itemize}
	
	For each simulated event, we compute the $\gamma$-ray flux by taking all relevant contributions for the SJ and CL models, as described above. Then, a simulated GRB is assumed to be detectable, if its prompt $\gamma$-ray flux exceeds the {\it Fermi} GBM threshold $F_{\rm lim}=5.8\times 10^{-7}\mbox{erg s}^{-1}\mbox{ cm}^{-2}$ \citep{Goldstein2017}; note that for sGRBs one can neglect the duration in the definition of the limiting flux.
	The detectability of GWs is a complex function of the viewing angle with respect to the merger plane \citep{Schutz2011,Allen2012,Hilborn2018}.
	Using the angular dependence of the detection probability given in \cite{Schutz2011} (see equation (27) therein) and scaling with the horizon distance for binary mergers of 220~Mpc, we define as detectable in GWs any simulated events that satisfy the condition:
	\begin{equation}
	\label{eq:GWdetect}
	\frac{220\, \mbox{Mpc}}{d}\bigg(\frac{1+6\cos^2\theta_{\rm obs}+\cos^4\theta_{\rm obs}}{8}\bigg)^{1/2}>1.
	\end{equation}
	The Monte Carlo process is repeated $10^5$ times, and the burst properties are recorded for events with joint $\gamma$-ray and GW detection or with just one of the two.
	
	\begin{figure*}
		\centering
		\includegraphics[width = .44\textwidth]{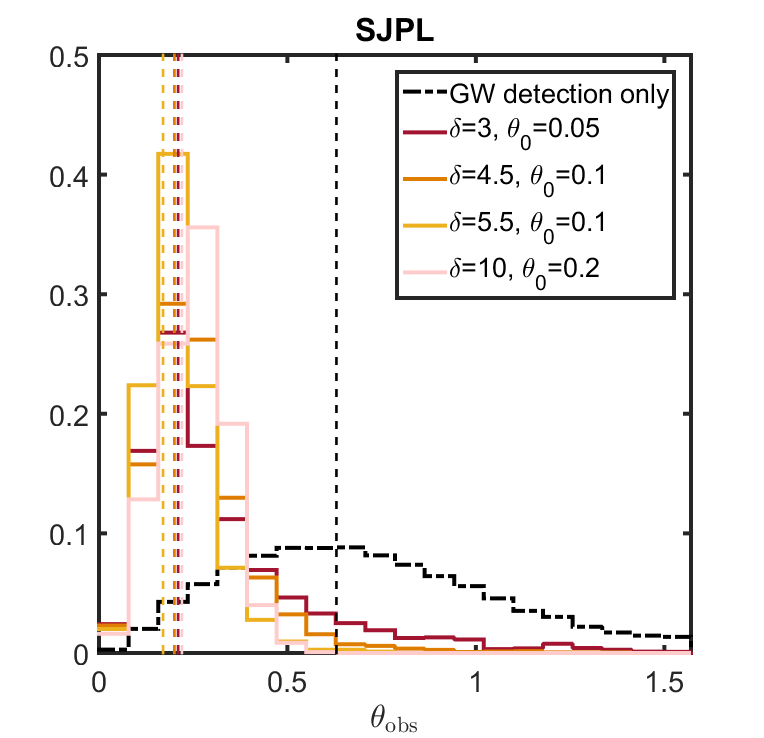}     
		\includegraphics[width = .44\textwidth]{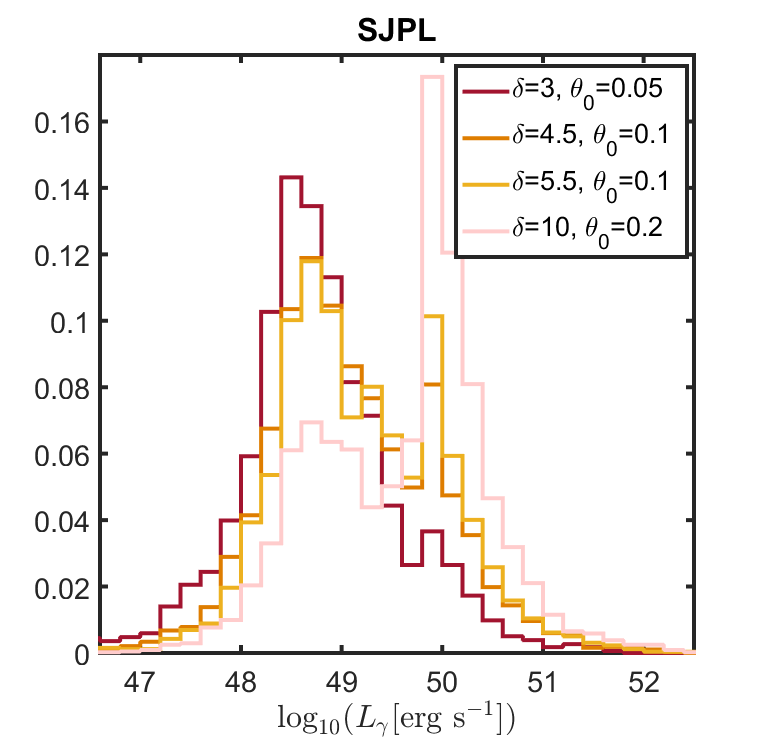}\\
		\includegraphics[width = .44\textwidth]{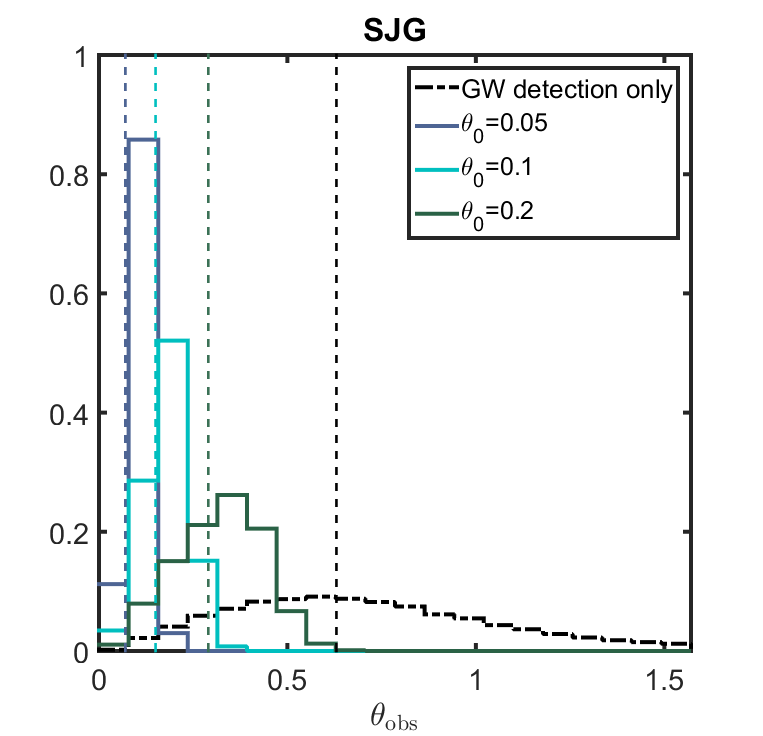}
		\includegraphics[width = .44\textwidth]{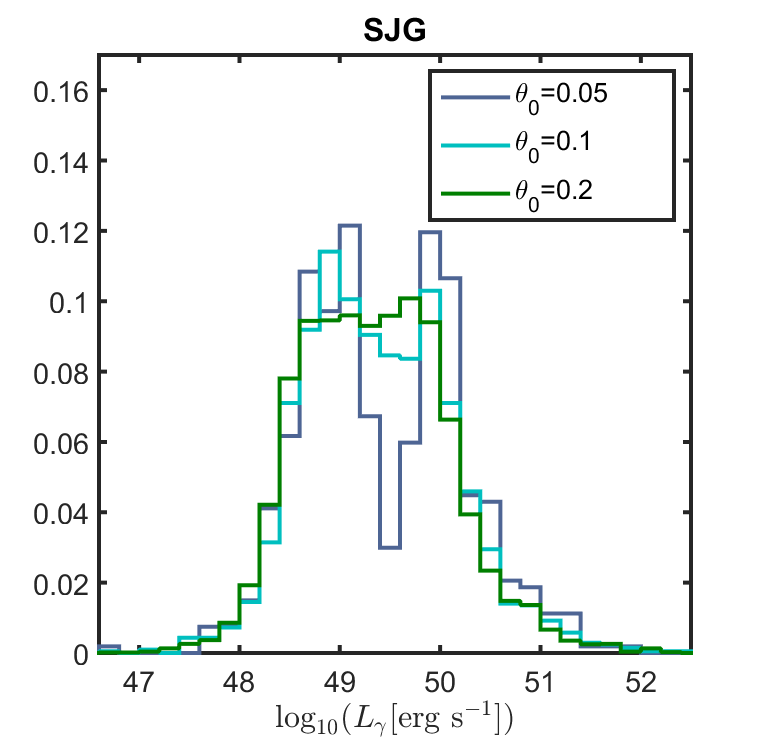}\\
		\includegraphics[width = .44\textwidth]{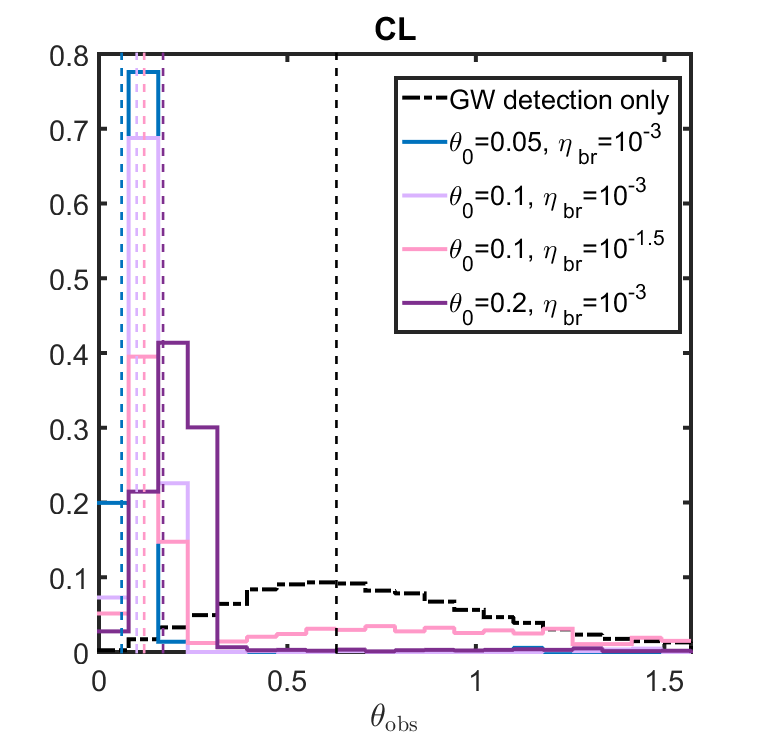}
		\includegraphics[width = .44\textwidth]{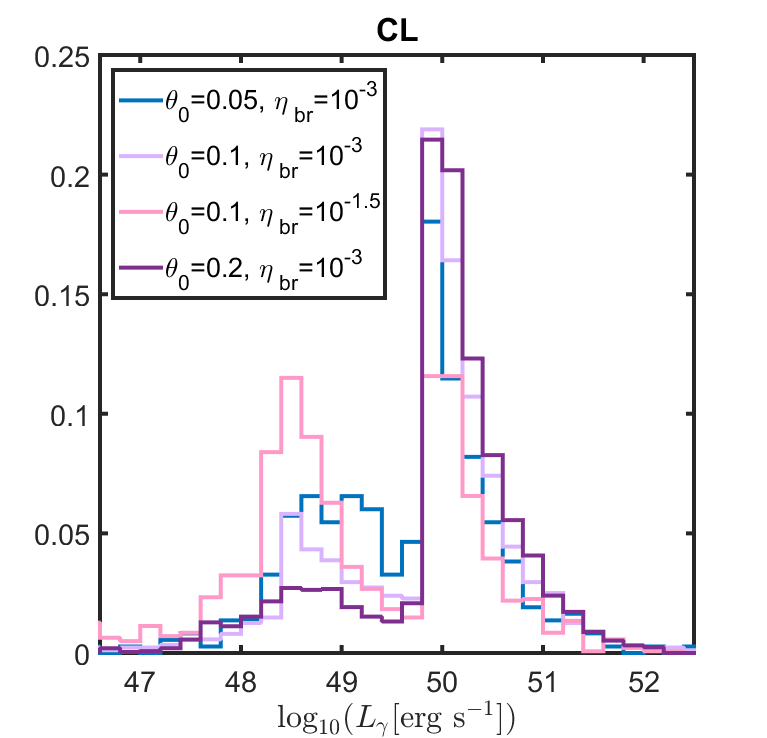}
		\caption{Probability distribution functions of observation angles (${\rm d}P/{\rm d}\theta_{\rm obs}$) and $\gamma$-ray (isotropic) luminosities (${\rm d}P/{\rm d}\log L_{\gamma}$) for different structure models discussed in \S\ref{sec:MC}. Results for a joint prompt GRB and GW detection are shown with solid lines in all panels. The dot-dashed histogram (left-hand side panels) shows the probability for a GW signal detection, as given by equation (\ref{eq:GWdetect}), without considering the prompt GRB emission. Dashed lines mark the median observation angle for different models, $\tilde{\theta}_{\rm obs}$. }
		\label{fig:distributions}
	\end{figure*} 
	\begin{table*}
		\begin{center}
			\caption{Detection probabilities of NS mergers obtained from our Monte Carlo simulations for different emission models (see \S\ref{sec:MC} for details). Values listed are calculated for $d_{\rm max} = 220$ Mpc  and 50 Mpc (in parenthesis). Note that all the probabilities listed here are conditional probabilities assuming that there is a GW detection.}
			\begin{threeparttable}
				\begin{tabular}{cccccc}\hline			 
					Model &  $\theta_0$ & $P_1$\tnote{(a)}& $P_2$\tnote{(b)}& $P_3$\tnote{(c)} & $\tilde{\theta}_{\rm obs}$\tnote{(d)} \\ \hline
					PL ($\delta=3$) & 0.05 & 0.94 (0.9)& 0.002 (0.001)& 0.05 (0.01)& 0.21 (0.5)\\
					PL ($\delta=4.5$) & 0.1 & 0.93 (0.93)& 0.006 (0.002)& 0.06 (0.07)& 0.2 (0.39)\\
					PL ($\delta=5.5$) & 0.1 & 0.95 (0.96)& 0.006 (0.002)& 0.04 (0.04)& 0.17 (0.28)\\
					PL ($\delta=10$) & 0.2 & 0.92 (0.96)& 0.03 (0.01)& 0.05 (0.03)& 0.22 (0.3)\\
					Gaussian & 0.05 & 0.98 (0.99)& 0.002 (0.001)& 0.02 (0.008)& 0.07 (0.19)\\
					Gaussian & 0.1 & 0.93 (0.96)& 0.008 (0.002)& 0.06 (0.03)& 0.15 (0.19)\\
					Gaussian & 0.2 & 0.76 (0.86)& 0.03 (0.01)& 0.2 (0.13)& 0.29 (0.38)\\
					CL ($\eta_{\rm br}\!=\!10^{-3}$) & 0.05 & 0.99 (0.99)& 0.002 (0.001)& 0.004 (0.003)& 0.06 (0.09)\\
					CL  ($\eta_{\rm br}\!=\!10^{-3}$) & 0.1 & 0.99 (0.99)& 0.006 (0.002)& 0.005 (0.008)& 0.1 (0.15)\\
					CL  ($\eta_{\rm br}\!=\!10^{-1.5}$) & 0.1 & 0.98 (0.96)& 0.006 (0.002)& 0.01 (0.03)& 0.12 (0.94)\\
					CL  ($\eta_{\rm br}\!=\!10^{-3}$) & 0.2 & 0.96 (0.97)& 0.03 (0.01)& 0.01 (0.02)& 0.17 (0.23)\\
					\hline  
					\label{tbl:prob}
				\end{tabular}
				\begin{tablenotes}
					\item[(a)] Probability of GW detection without an accompanying $\gamma$-ray signal.
					\item[(b)] Probability of detection of an on-axis sGRB given a GW detection.
					\item [(c)] Probability of detection of an off-axis sGRB given a GW detection.
					\item [(d)] Median value computed from joint $\gamma$-ray  and GW detections.
				\end{tablenotes}
			\end{threeparttable}
		\end{center}
	\end{table*}   
	\subsection{Results}
	The distributions of observation angles and prompt luminosities of GRBs with joint $\gamma$-ray and GW detections are presented respectively in the left and right panels of Fig.~\ref{fig:distributions} (solid lines). The dot-dashed histogram (left-hand side panels) shows the probability that a GW signal is detected for a given observation angle, as given by equation (\ref{eq:GWdetect}), i.e., independent of the detectability of the prompt GRB emission. The typical observation angle is slightly larger than $\theta_0$, with the median values, $\tilde{\theta}_{\rm obs}$, typically being in the range $0.1-0.25$ and approaching $\theta_0$ as the structure becomes steeper. This conclusion remains true even for the cocoon models explored here, due to their intrinsically weak emission. Only very radiatively efficient cocoon models, for which the cocoon is seen much more often than the on-axis GRB, may be seen much further off-axis. For example, for $\theta_0=0.1$ one would require $\eta_{\rm br}\ge 0.1$ in order to obtain $\tilde{\theta}_{\rm obs}\gtrsim 0.4$. As a comparison, the typical observation angle of a GW detection alone is $\tilde{\theta}_{\rm obs}\approx 0.65$ (see black vertical lines in the left-hand side panels of Fig.~\ref{fig:distributions}). 
	
	The luminosity distribution is typically double-peaked, with a high-luminosity peak corresponding to jets seen on-axis (peaking at $\approx 2L_{\rm min}$) and a low-luminosity peak corresponding mostly to off-axis jets (peaking at $\approx 2 L_{\rm min}(\tilde{\theta}_{\rm obs}/\theta_0)^{-\delta}$, $ 2 L_{\rm min}\exp[-(\tilde{\theta}_{\rm obs}/\theta_0)^2]$ for the SJPL and SJG models respectively. A smaller low luminosity peak is seen also for the CL model, this corresponds to the off line-of-sight contribution discussed in \S \ref{sec:MC}. The location of this peak in the figure is dominated by the $\gamma$-ray detectability threshold: $\approx2\times 4\pi \tilde{r}^2F_{\rm lim}$ where $\tilde{r}$ is the median distance of jointly detected events. A small contribution to the low-luminosity peak in the CL model, comes from the break-out luminosity of failed jets. The bi-modality is naturally more prominent for steeper structure models (see e.g. top right-hand side panel in Fig.~\ref{fig:distributions}) and becomes apparent when the two peaks discussed above are sufficiently separated. Gaussian structured-jet models produce symmetric bi-modal luminosity distributions for $\theta_0 \lesssim 0.2$, while the two peaks merge into one for larger $\theta_0$ values (see middle right-hand side panel).

	It is constructive to consider the likelihood of the three following outcomes: (i) no sGRB detection given a GW detection, (ii) a joint GW and a ``regular" sGRB detection ($\theta_{\rm obs}\lesssim \theta_0$), and (iii) a joint GW and misaligned sGRB  detection ($\theta_{\rm obs}>\theta_0$). The probabilities for the different models considered above and the median observation angles, given a joint detection, are presented in Table \ref{tbl:prob}. Excluding the SJG model with $\theta_0=0.2$ that is in contention with our results in \S \ref{sec:angles} and with limits from GRB170817, the probability for a detection of GW from a NS merger without an accompanying $\gamma$-ray signal is very large in all models considered here (i.e., $\approx 90-99\%$). In the more rare case of a joint detection, the sGRB is typically 1 to 10  times more likely to be seen off-axis.
	
	We explored also the effect of the limiting distance $d_{\max}$ on the aforementioned probabilities. In most simulations we performed, we took as a canonical value $d_{\rm max}=220$~Mpc, which as described above is the maximum distance to which NS mergers can be detected with near future GW detectors. A smaller limiting distance ( $d_{\rm max}=50$~Mpc) has an interesting effect on the detectability of events (see Table~\ref{tbl:prob}). While the fraction of events that can be detected electromagnetically increases as $d_{\rm max}^{-2}$, the probability of a sGRB detection given a GW observation does not change significantly by switching $d_{\rm max}$ from 220~Mpc to 50~Mpc. This can be understood as follows. The detectability of GWs becomes 100$\%$ for events with $d<50$~Mpc and the typical observation angle of a GW-detected event increases as $d_{\rm max}$ decreases. Consequently, the probability of observing an on-axis sGRB actually goes down for smaller $d_{\rm max}$. Similar conclusions hold for any sGRB (not only on-axis), if the angular profile is very steep or its isotropic component is very dim.

	GRB170817 could have been detected in $\gamma$-rays up to a distance of $50$~Mpc \citep{Goldstein2017}, but instead it occurred at a distance of $40$~Mpc. Moreover, the large inferred isotropic equivalent core luminosity in GRB170817 \citep{Mooley2018}, which is roughly two orders of magnitude above those of typical sGRBs, suggests an accordingly luminous off-axis $\gamma$-ray emission.
	With a detection horizon of 220~Mpc, only a small fraction of future events are expected to exhibit similar properties to those of GRB170817. In models with a strong angular structure (e.g., SJPL models), bursts must be observed at small enough $\theta_{\rm obs}$, so that there is still a significant amount of power in the material traveling towards the observer. This leads to a further reduction in the expected rate of future GRB170817 - like events.
	
\begin{figure}
		\centering
		\includegraphics[width = .4\textwidth]{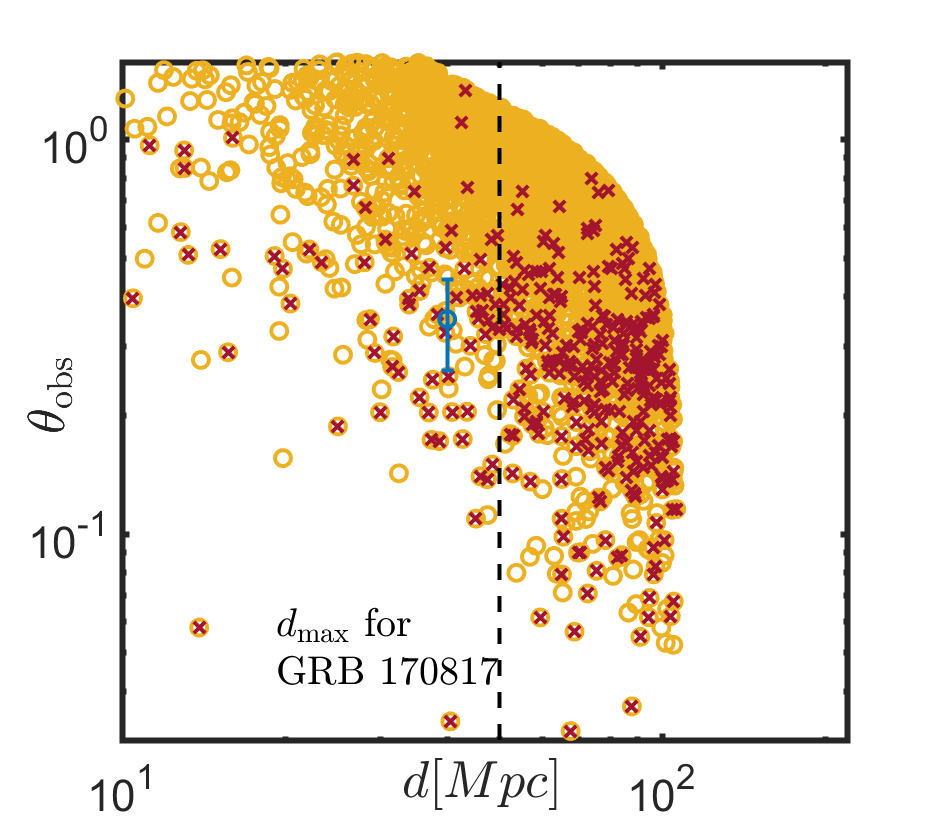}
		\includegraphics[width = .4\textwidth]{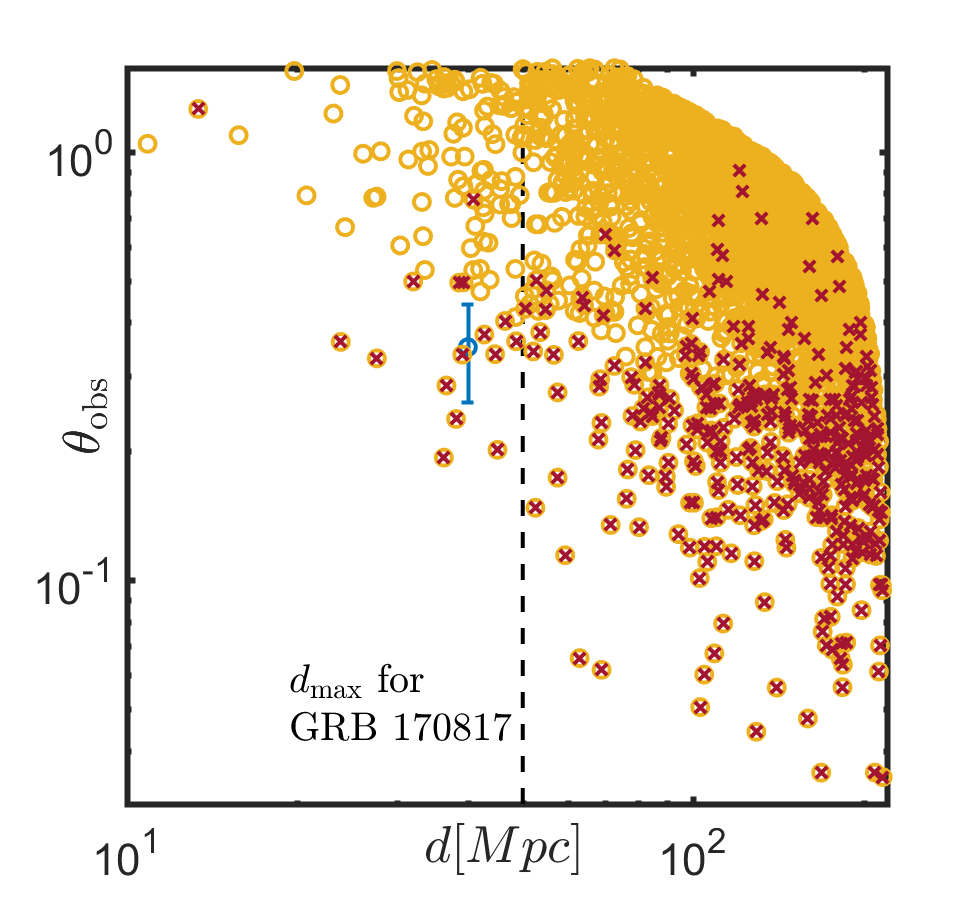}
		\caption{Distances and observation angles for mergers detected in GW alone (yellow circles) and for mergers with joint GW and $\gamma$-ray detections (red crosses). We present the distribution with a detection horizon of 107Mpc (top), corresponding to the effective detection horizon for GW170817 or 220Mpc (bottom), suitable for future events. The inferred values for GRB170817 are shown with a blue symbol. The maximum distance at which it could have been observed in $\gamma$-rays is depicted by the vertical dashed line. Results are shown for a SJPL model with $\theta_0=0.1$ and $\delta=4.5$.}
		\label{fig:thetadistmax}
	\end{figure}

\begin{figure*}
		\centering
		\includegraphics[width = .49\textwidth]{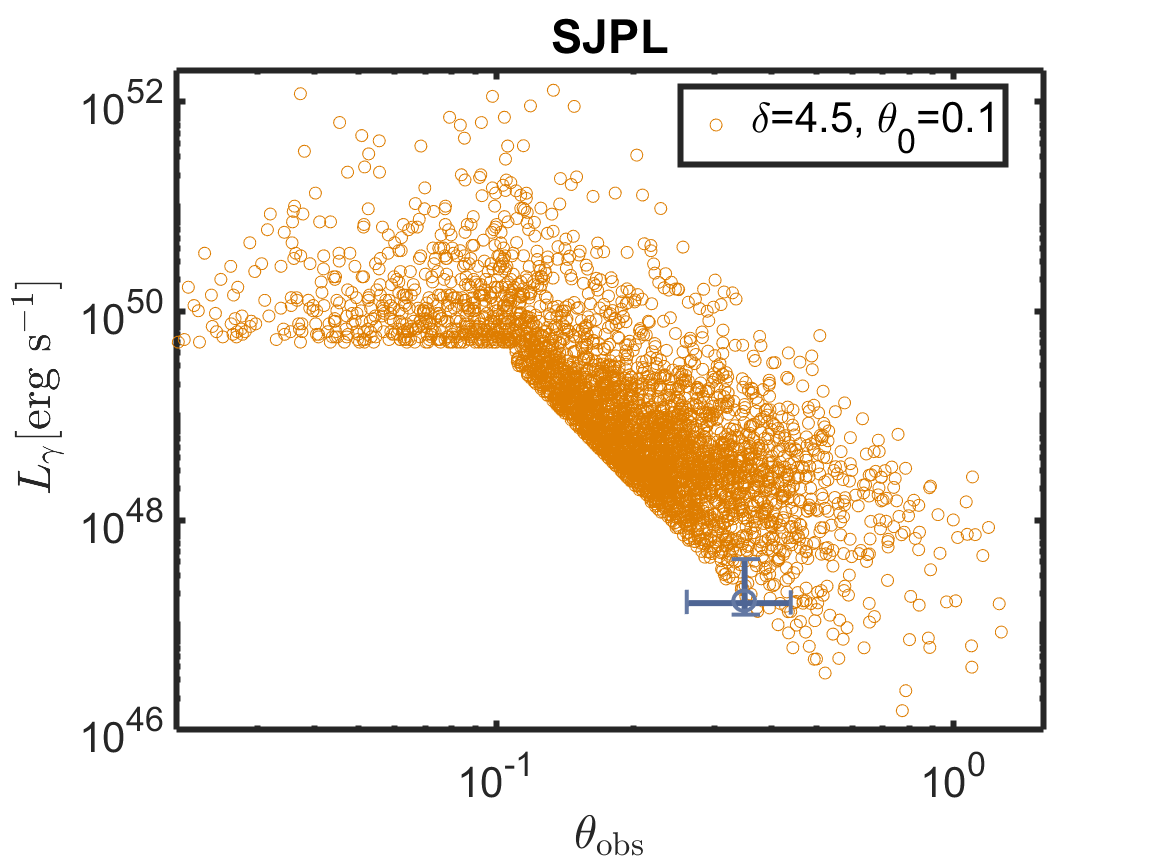}\
		\includegraphics[width = .49\textwidth]{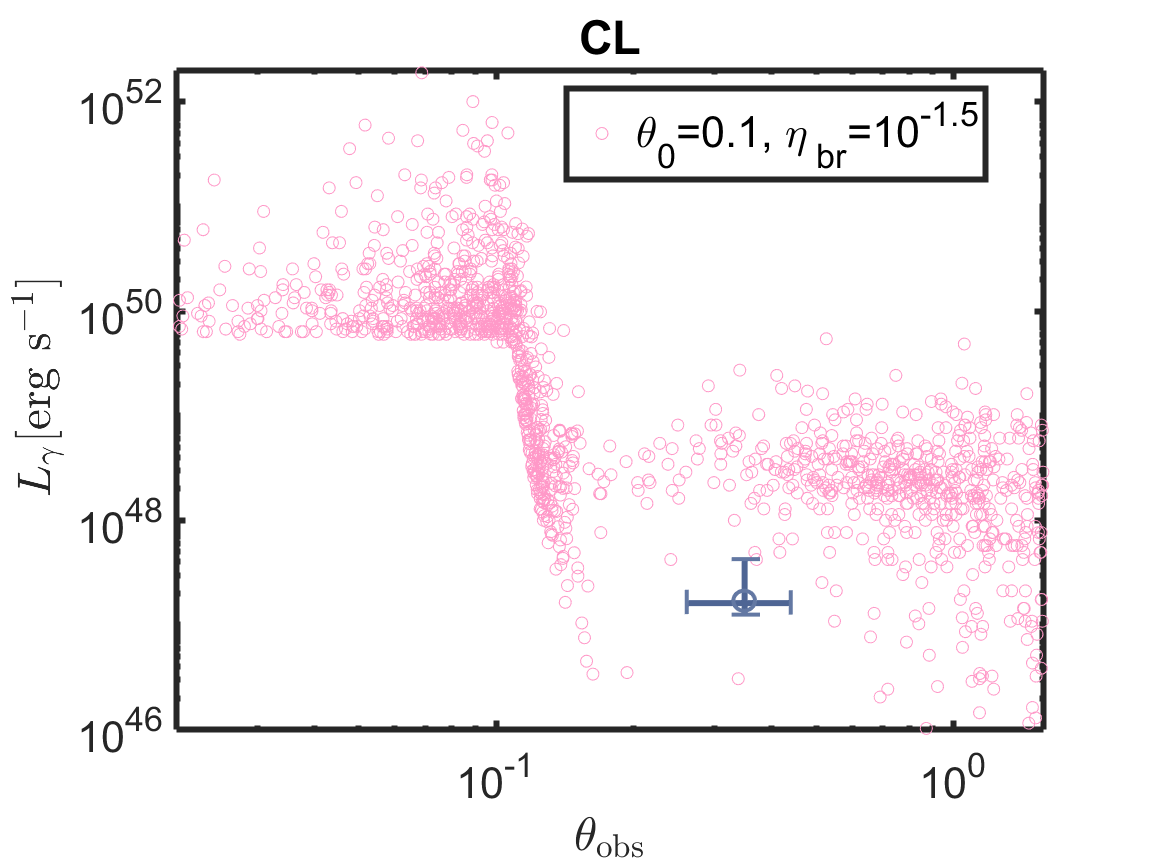}
		\caption{Two-dimensional distributions of observation angles and $\gamma$-ray luminosities for different structure models assuming a joint prompt GRB and GW detection. The location of GRB170817 on these plots is marked with blue crosses that denote the 1$\sigma$ error in luminosity \citep{Zhang2018} and observation angle \citep{Mooley2018}.}
		\label{fig:jointdist}
	\end{figure*}

	The effective LIGO horizon distance for the detection of GW170817 was 107 Mpc \citep{GW170817}. Given a GW detection alone, we find that the probability of an event having $\theta_{\rm obs}\leq 0.44$ (as inferred for GRB170817; see \citealt{Mooley2018}) and occurring at a distance of $d\leq 50$Mpc is $\approx 4$\% (see also \citealt{Lamb2017}). This is demonstrated in figure~\ref{fig:thetadistmax} where we plot the observation angles and distances of events with joint $\gamma$-ray and GW detections (red crosses) from a simulation of SJPL model with $\theta_0=0.1$ and $\delta=4.5$. Top and bottom panels show results for an effective GW horizon distance of 107~Mpc and 220~Mpc,  respectively. The median distance of future joint detection events is $\approx 150$~Mpc and correspondingly the median GRB luminosity is $7\times 10^{48}\mbox{erg s}^{-1} \gg L_{\rm GRB170817}$. The fact that GRB170817 is an atypical event of the sample of expected joint prompt and GW detections is also illustrated in figure~\ref{fig:jointdist}, where we present the expected $L_{\gamma}-\theta_{\rm obs}$ distributions of joint detections for two models of prompt emission given a detection horizon of 220Mpc. For both models, the properties of GRB170817 (blue crosses) are not representative of the sample. Only $\sim 0.16\%$ and $2\%$ of the simulated events (jointly detected in $\gamma$-rays and GWs) in the CL and SJPL models, respectively, is found to have $0.25\le \theta_{\rm obs}\le 0.45$ and $5\times10^{46}$~erg s$^{-1}$$\le L_{\gamma}\le 4.8\times10^{47}$~erg s$^{-1}$. The shape of the observed two-dimensional distributions is driven by the emission model, since the emission is typically dominated by material that is radiating within its $1/\Gamma$ beaming cone from the observer. A differentiation between the underlying emission models would require a sufficiently large number of joint GW and GRB detections (see also \S\ref{sec:discussion}).
	
	\section{The cocoon emission}
	\label{sec:cocooncooling}
	We estimate here the thermal energy stored in the cocoon, the fraction of it radiated in $\gamma$-rays at breakout and the cooling emission signature arising when the thermal cocoon eventually becomes optically transparent. The latter corresponds to a $\sim$ hours long UV signal, that may be detectable, provided sufficiently rapid localization of the NS merger.

	The peak thermal energy stored in the cocoon (as a function of time) is found by \cite{Duffell2018}. As mentioned above, here we assume that the that the injected opening angle is equal to the opening angle after breakout. Re-writing equations 10, 21 in \cite{Duffell2018} in a slightly different form, we obtain:
	\begin{equation}
	\label{eq:peakEth}
	E_{\rm Th}= \left \{ 
	\begin{array}{ll}
	0.17 E_j & E_{\rm cr}\le E_j \le 30 E_{\rm cr}  \\
	1.14E_{\rm cr} &  E_j > 30 E_{\rm cr},
	\end{array}
	\right.
	\end{equation}
	where $E_{\rm cr}\approx 5\times 10^{49}\,\theta_0^2$~erg and $E_j$ is the beaming corrected kinetic energy of the jet, $E_j\approx 0.5 \, \theta_0^2 L_0 t_e /\eta_{\gamma}$. The peak thermal energy in equation (\ref{eq:peakEth}) is reached at a time
	\begin{equation}
	\label{eq:peakTth}
	t_{\rm peak}= \left \{ 
	\begin{array}{ll}
	2t_e (<t_{\rm br}) & E_{\rm cr}\le E_j \le 30 E_{\rm cr}  \\
	t_{\rm br}&  E_j > 30 E_{\rm cr},
	\end{array}
	\right.
	\end{equation}
	where $t_{\rm br}=3t_e E_{\rm cr}/E_j$ is the time it takes the jet to breakout from the ejecta. The thermal energy stored in the cocoon provides a strict upper limit on its radiated energy and luminosity. This is because the cocoon is expected to  be initially optically thick, in which case only a small portion of the thermal energy, can escape to the observer (see \citealt{SN2010} and references therein). For GRB170817, the latest modelling efforts suggest a total (beaming corrected) energy of $10^{50}$~erg \citep{Mooley2018} which corresponds to  $E_{\rm Th}\approx 3\times 10^{49}$~erg. At the same time, cocoon breakout models for the prompt $\gamma$-rays radiate only $E_{\rm br}=4\times 10^{46}$~erg at breakout \citep{Gottlieb2017,Lazzati2017}, implying an efficiency of $\eta_{\rm br}\equiv E_{\rm br}/E_{\rm Th}\sim 10^{-3}$ for this phase. In what follows we adopt $\eta_{\rm br}=10^{-3}$ as a canonical value.

	The entire thermal energy is eventually released on the time-scale for which the medium reaches transparency $t_{\rm thin}$. As an indicative example, let us consider a characteristic velocity $v=0.5c$ for the cocoon ejecta of mass $M_{\rm co}$. Assuming that the kinetic energy of the cocoon is comparable to its thermal energy \citep{Duffell2018} ($E_{\rm Th}=0.5M_{\rm co}v^2$), $t_{\rm thin}$ is estimated as:
	\begin{eqnarray}
	\label{eq:tthin}
	&t_{\rm thin}=\sqrt{\frac{\kappa M_{\rm co}}{4\pi c v}}=\sqrt{\frac{2E_{\rm Th}\kappa }{4\pi c v^3}}= \\& \left \{ 
	\begin{array}{ll}
	1600 E_{50}^{1/ 2}v_{.5}^{-3/2}\kappa_{0.1}^{1/2} \mbox{ s } & E_{\rm cr}\le E_j \le 30 E_{\rm cr} \\
	300 v_{.5}^{-3/2}\theta_{0,.1}\kappa_{0.1}^{1/2} \mbox{ s } &  E_j > 30 E_{\rm cr},\nonumber
	\end{array}
	\right.
	\end{eqnarray}
	where $E_{50}\equiv E_j/10^{50}\mbox{ erg},v_{.5}\equiv c/0.5c,\theta_{0,.1}\equiv \theta_0/0.1$, and $\kappa_{0.1}=\kappa/0.1\mbox{cm}^2\mbox{g}^{-1}$ is the Thomson opacity (cross section per unit mass). For the adopted parameter values, we find that $t_{\rm thin} \gg t_{\rm br} \gtrsim t_{\rm peak}$ (see equation \ref{eq:peakTth}). This is consistent with the small efficiencies for the cocoon breakout luminosity, $\eta_{\rm br}\ll 1$ inferred for GW170817, as discussed above. The peak (isotropic equivalent) cocoon breakout luminosity can be approximated by using equations (\ref{eq:peakEth}) and (\ref{eq:peakTth}):
	\begin{equation}
	\label{Lgco}
	L_{\gamma,\rm co}\approx \eta_{\rm br}\frac{E_{\rm Th}}{t_{\rm peak}}\approx \frac{\theta_0^2}{2} \frac{\eta_{\rm br}}{\eta_{\gamma}}k L_0,
	\end{equation}
	where $k$ is a numerical constant that is $k\approx 0.08$ for $E_{\rm cr}\le E_j \le 30 E_{\rm cr}$ and $k\approx 0.38$ for $E_j > 30 E_{\rm cr}$. The spectrum of the cocoon is expected to be quasi-thermal (possibly following a Wien spectrum, see \citealt{Gottlieb2017}) with a characteristic temperature $k_BT=50\,\Gamma_{br}$keV (where $\Gamma_{br}\approx few$ is the Lorentz factor of the material at the break-out shell). As more and more matter becomes transparent, the prompt luminosity released at breakout, evolves smoothly to the cooling signal described below. The exact temporal evolution depends on the velocity distribution of the material behind the break-out shell. Using the parameterization adopted above, we get $L\propto \eta(t)t^{-2}$ (this can be shown by comparing equation \ref{Lgco} with equation \ref{eq:Lcool} below) where $\eta(t_{\rm peak})\equiv \eta_{br}\ll1$ and at later times $\eta(t)$ depends on the shock profile (eventually reaching $\eta(t_{\rm thin})=1$).
	
	The cocoon ejecta cools due to adiabatic energy losses until it becomes optically thin and the remaining energy $E_{\rm cool}< E_{\rm Th}$ is radiated away. This will happen when the time-scale for the diffusion of photons from the cocoon becomes comparable to the dynamical timescale. At this point, the cocoon radius is $vt_{\rm thin}$ and the energy $E_{\rm cool}$ is given by:
	\begin{eqnarray}
	& E_{\rm cool}=E_{\rm Th}\frac{t_{\rm peak}}{t_{\rm thin}}=\\&\left \{ 
	\begin{array}{ll}\!\!
	6\times 10^{45}E_{50}^{1/2}v_{.5}^{3/ 2}t_{e,.3}\kappa_{0.1}^{-1/2}\mbox{ erg }& E_{\rm cr}\le E_j \le 30 E_{\rm cr} \\
	\!\!9\!\times \! 10^{42} E_{50}^{-1}v_{.5}^{3/ 2}\theta_{0,.1}^3t_{e,.3}\kappa_{0.1}^{-1/2} \mbox{ erg}  &  E_j > 30 E_{\rm cr},\nonumber
	\end{array}
	\right.
	\end{eqnarray}
	where $t_{e,.3}\equiv t_e/0.3\mbox{ s}$.
	This energy is radiated away on a time-scale of $t_{\rm thin}$. The cooling luminosity associated with this phase is then given by:
	\begin{eqnarray}
	\label{eq:Lcool}
	& L_{\rm cool}=\frac{E_{\rm cool}}{t_{\rm thin}}=\\&\left \{ 
	\begin{array}{ll}\!\!
	4\times 10^{42}v_{.5}^{3}t_{e,.3}\kappa_{0.1}^{-1}\mbox{ erg s}^{-1}& E_{\rm cr}\le E_j \le 30 E_{\rm cr} \\
	\!\!3\!\times \! 10^{40} E_{50}^{-1}v_{.5}^{3}\theta_{0,.1}^2t_{e,.3}\kappa_{0.1}^{-1} \mbox{ erg s}^{-1}  &  E_j > 30 E_{\rm cr}.\nonumber
	\end{array}
	\right.
	\end{eqnarray}
	Notice that the cooling luminosity is very sensitive to the velocity of the outflow ($L_{\rm cool}\propto v^3$). Since this radiation is fully thermalized, and originates from a radius of $vt_{\rm thin}$, we can estimate its black body temperature as:
	\begin{eqnarray}
	& T_{\rm Th}=\bigg(\frac{L_{\rm cool}}{4\pi \sigma v^2 t_{\rm thin}^2}\bigg)^{1/4}=\\&\left \{ 
	\begin{array}{ll}
	5\times 10^{4}E_{50}^{-1/4}v_{.5}t_{e,.3}^{1/4}\kappa_{0.1}^{-1/2}\mbox{ K }& E_{\rm cr}\le E_j \le 30 E_{\rm cr} \\
	4\times \! 10^{4} E_{50}^{-1/4}v_{.5}t_{e,.3}^{1/4}\kappa_{0.1}^{-1/2} \mbox{ K }  &  E_j > 30 E_{\rm cr}.\nonumber
	\end{array}
	\right.
	\end{eqnarray}
	Therefore we expect the cooling emission from the cocoon to peak at the UV band (see also \citealt{NP2017}). For a limiting {\it Swift}-UVOT B band flux of $\sim 5\times 10^{-3}$mJy in 1000~s \citep{Gehrels2004}, and $v=0.5c$ the derived luminosities from equation (\ref{eq:Lcool}) imply that this emission should be detectable for future NS merger events up to $\sim 900$~Mpc if $E_{\rm cr}\le E_j \le 30 E_{\rm cr}$ and up to $\sim 80$~Mpc if $E_j > 30 E_{\rm cr}$. This of course, is provided that these events can be localized rapidly enough, in time to catch this signal (see also equation (\ref{eq:tthin})).
	This can be a  challenging prospect. As an example, the EM counterpart of GW170817, was only first detected $0.5$ days after the GW detection \citep{Coulter2017}.
	
	\begin{table}
		\begin{center}
			\caption{Expected number of NS mergers detectable in GWs in the case of: no {\it Fermi}/GBM detection, an on-axis GRB detection, and an off-axis GRB detection. The numbers (with their 1$\sigma$ statistical errors \citep{Gehrels1986}) are obtained from our simulations for a limiting distance of 220~Mpc and a ten-year period, assuming a merger rate $R_{\rm merg}=1540$~Gpc$^{-3}$~yr$^{-1}$.  For the assumed parameters, $\sim 192$ NS mergers are expected.}
			\resizebox{0.4848\textwidth}{!}{
				\begin{tabular}{llccc}\hline	
					Model & parameters & {\it Fermi} undetected & on-axis & off-axis  \\ 
					\hline
					PL & $\delta=3$, $\theta_0=0.05$ &  $180.8\pm13.4$ & $<1.8$ & $9.6^{+3.5}_{-3.5}$\\  [5pt]
					PL & $\delta=4.5$, $\theta_0=0.1$& $178.9\pm13.4$ & $1.1^{+2.1}_{-1.0}$ & $11.5^{+3.9}_{-3.8}$\\  	[5pt]				            	PL & $\delta=5.5$, $\theta_0=0.1$ & $182.7\pm13.5$ & $1.1^{+2.1}_{-1.0}$ & $7.7^{+3.1}_{-3.3}$\\ [5pt]
					PL & $\delta=10$, $\theta_0=0.1$ & $177.0\pm13.3$ & $5.8^{+2.6}_{-2.9}$ & $9.6^{+3.5}_{-3.5}$\\[5pt]
					GS & $\theta_0=0.05$ &  $188.5\pm13.7$ & $<1.8$ & $3.8^{+2.1}_{-2.5}$\\[5pt]
					GS & $\theta_0=0.1$ & $178.9\pm13.4$ & $1.5^{+1.8}_{-1.4}$ & $11.5^{+3.9}_{-3.8}$\\[5pt]
					GS & $\theta_0=0.2$ & $146.2\pm12.2$ & $5.8^{+2.6}_{-2.9}$& $38.5^{+6.7}_{-6.6}$\\[5pt]
					CL & $\eta_{\rm br}\!=\!10^{-3}$, $\theta_0=0.05$ & $190.4\pm13.8$ & $<1.8$ & $<1.8$\\[5pt]
					CL & $\eta_{\rm br}\!=\!10^{-3}$, $\theta_0=0.1$ &$190.4\pm13.8$& $1.1^{+2.1}_{-1.0}$& $1.0^{+2.3}_{-0.8}$\\[5pt]
					CL & $\eta_{\rm br}\!=\!10^{-3/2}$, $\theta_0=0.1$ & $188.5\pm13.7$& $1.1^{+2.1}_{-1.0}$&$1.9^{+1.4}_{-1.7}$ \\[5pt]
					CL & $\eta_{\rm br}\!=\!10^{-3}$, $\theta_0=0.2$ &  $184.6\pm13.6$ & $5.8^{+2.6}_{-2.9}$&$1.9^{+1.4}_{-1.7}$\\
					\hline  
					\label{tbl:number}
				\end{tabular}
			}
		\end{center}
	\end{table}    
	
	\section{Discussion}
	\label{sec:discussion}
	The first discovery of GW from a NS merger allowed us to significantly improve our understanding of sGRB jets.
	Assuming all sGRBs arise from NS-NS (or NS-BH) mergers, the intrinsic rate of sGRBs should be at most comparable to the merger rate inferred by advanced LIGO/Virgo. This implies that if sGRB jets have a universal, luminosity independent structure, their typical opening angles should be $\theta_0\gtrsim 0.07$ (see \S \ref{sec:angles}). At the same time, the observed population of sGRBs with {\it Swift} in the last 14 years places a lower limit of $\sim 250$~Mpc on the distance from which we have observed an on-axis sGRB. This leads to an upper limit on the rate of local on-axis sGRBs and therefore on their typical opening angles $\theta_0\lesssim 0.1$. These limits on the opening angle are consistent with values inferred for GRB170817 from afterglow modelling and from the measurements of superluminal motion. Furthermore, they imply that the NS merger rate is comparable to that of sGRBs and reveals that a fraction of order unity of NS mergers must lead to sGRBs. In other words, sGRB jets typically manage to breakout of the NS-merger ejecta, in contrast to collapsar GRB jets.
	
	The large fraction of successful sGRB jets and the typical opening angles of $ \theta_0 \sim 0.1$ are consistent with a critical breakout luminosity (estimated from hydrodynamical simulations of the interaction between the sGRB jet and the NS merger ejecta) being close to the ``canonically" assumed minimal luminosity of the sGRB luminosity function, namely $L_{\rm min}\approx 5\times 10^{49}\, \mbox{erg s}^{-1}$. This consideration demonstrates that the role of failed jets in shaping the observed luminosity function of sGRBs cannot be significant (see \S \ref{sec:failedjets}). At the same time, we have also shown here that the angular structure of sGRBs is challenged to reproduce the observed luminosity function, given that the required structure is very shallow in contention with observational indications from GRB170817 (see \S \ref{sec:angular}). If this is the case, the implication is that the broken power-law luminosity function of sGRBs has an intrinsic origin and that the inferred break of the luminosity function, at an isotropic $\gamma$-ray luminosity of $L_*\approx 2\times 10^{52}\, \mbox{erg s}^{-1}$ (corresponding to a beaming corrected jet mechanical power of $\sim 10^{51}\mbox{erg s}^{-1}$) reveals an intrinsic characteristic luminosity of sGRB jets. One possible interpretation that holds for magnetic jets, powered by the Blandford Znajek mechanism \citep{BZ1977}, is that the value of $L_*$ reflects a characteristic accretion rate, below which the accretion disk is advection dominated (ADAF), and above which it becomes dominated by neutrino cooling (NDAF) \citep{Giannios2007}. Indeed, \cite{Kawanaka2013} have shown that a sharp change in the jet power may occur due to this transition at accretion rates $\dot{M}\approx 0.003-0.01M_{\odot}\,\mbox{s}^{-1}$, consistent with the beaming corrected jet luminosity mentioned above, assuming $L\approx 0.1\dot{M} c^2$.
	
	While the joint discovery of  GW170817 and GRB170817 has helped to constrain the fraction of successful GRB jets, their opening angles, and their luminosity function as discussed above, the angular structure of GRB jets and the nature of their $\gamma$-ray emission are still uncertain. In this paper, we considered different off-axis emission models, motivated by the analysis above and by observations of GRB170817. We showed that $90\%-99\%$ of future GW events accompanied by a successful GRB jet and detected up to a distance of $220$~Mpc, should not be accompanied by any detectable prompt GRB signal. In the comparatively rare cases of a joint GRB and GW detections, we find that for each GRB observed on-axis $\sim 1-10$  GRBs should be observed at angles beyond the jet core. 
	
	The distribution of prompt luminosities and observation angles from joint GRB and GW detections can help to distinguish between off-axis prompt emission models (see figure~\ref{fig:jointdist}). For example, let us assume a NS merger rate of $R_{\rm merg}=1540 \, \mbox{Gpc}^{-3} \mbox{yr}^{-1}$, a ratio of failed to successful GRB jets $r_{\rm fail}=1$, and $\theta_0=0.1$. Then, angular structure models with $L(\theta>\theta_0)\propto \theta^{-\delta}$ and  $\delta=4.5$ lead to $\sim 19.2$ GW detectable mergers per year (up to 220~Mpc), out of which $\sim18$ without {\it Fermi}/GBM $\gamma$-ray detection, 0.1 with an on-axis GRB detection, and 1.1 with an off-axis GRB detection. Alternatively, for cocoon models with a breakout efficiency of $\eta_{\rm br}=10^{-3}$ (and $\theta_0=0.1$), we have $\sim 19$ events per year with no accompanying GRB, and 0.1 with on-axis or off-axis GRBs. Table~\ref{tbl:number} summarizes the expected number of events within a period of ten years for the emission models discussed in \S \ref{sec:MC}. Inspection of the table shows that the detection rates of off-axis GRBs accompanying GW-detected mergers are the key for differentiating between the models. For example, the lack of any other prompt GRB detections from NS mergers within the next 10 years (for the assumed parameters), would be in strong tension with the predictions of the structured jet model. The thermal energy in the cocoon may also be observed via its cooling emission that is expected to lead to a UV thermal signal at $\sim 10^3$~s after the trigger (see equations (\ref{eq:tthin}) and (\ref{eq:Lcool})). Provided that the NS merger can be located rapidly enough, the cooling emission of the cocoon may be detectable up to a distance of $\sim 900$~Mpc. It turns out that events similar to GRB170817 are rare. This is a combination of the small distance and observation angle of GRB170817 and the high inferred luminosity at the core of that event compared to typical sGRBs.

	In conclusion, the discovery of the first GRB associated with a NS merger, GRB170817, has already significantly improved our knowledge of short GRB jets. Nonetheless, the nature of the prompt signal that is seen by observers far from the jet cores remains unclear.
	The association of a $\gamma$-ray signal with a GW event could allow us to detect orders of magnitude fainter signals associated with GRBs, which would otherwise be undetected for cosmological events \citep[see also][]{Beniamini2018}. This, coupled with the relatively large detection rate of NS mergers as inferred from GW170817, implies that the existing models could be differentiated observationally within the next several years, opening up the window towards a more detailed understanding and future studies of short GRB jets. 
	\section*{Acknowledgments}
	The authors would like to thank the referee for a constructive report, and Tsvi Piran, Ehud Nakar, Pawan Kumar, Omer Bromberg and Kenta Hotokezaka for helpful discussions. RBD thanks Danielle Taylor for useful discussions. MP acknowledges support from the Lyman Jr.~Spitzer Postdoctoral Fellowship and NASA grant 80NSSC18K1745. DG acknowledges support from NASA grants NNX16AB32G and NNX17AG21G. RBD and DG acknowledge support from the National Science Foundation under Grants 1816694 and 1816136.


	\bsp	
	\label{lastpage}

\begin{thebibliography}{91}
	\expandafter\ifx\csname natexlab\endcsname\relax\def\natexlab#1{#1}\fi
	
	\bibitem[{{Abbott} {et~al}\mbox{.}(2017{\natexlab{a}}){Abbott}, {Abbott},
		{Abbott}, {Acernese}, {Ackley}, {Adams}, {Adams}, {Addesso}, {Adhikari},
		{Adya}, \& et~al.}]{joint170817}
	{Abbott} B.~P. {et~al.}, 2017{\natexlab{a}}, \apjl, 848, L13
	
	\bibitem[{{Abbott} {et~al}\mbox{.}(2017{\natexlab{b}}){Abbott}, {Abbott},
		{Abbott}, {Acernese}, {Ackley}, {Adams}, {Adams}, {Addesso}, {Adhikari},
		{Adya}, \& et~al.}]{GW170817}
	{Abbott} B.~P. {et~al.}, 2017{\natexlab{b}}, Physical Review Letters, 119,
	161101
	
	\bibitem[{{Allen} {et~al}\mbox{.}(2012){Allen}, {Anderson}, {Brady}, {Brown},
		\& {Creighton}}]{Allen2012}
	{Allen} B., {Anderson} W.~G., {Brady} P.~R., {Brown} D.~A., {Creighton}
	J.~D.~E., 2012, \prd, 85, 122006
	
	\bibitem[{{Aloy}, {Janka} \& {M{\"u}ller}(2005){Aloy}, {Janka}, \&
		{M{\"u}ller}}]{Aloy2005}
	{Aloy} M.~A., {Janka} H.-T., {M{\"u}ller} E., 2005, \aap, 436, 273
	
	\bibitem[{{Band} {et~al}\mbox{.}(1993){Band}, {Matteson}, {Ford}, {Schaefer},
		{Palmer}, {Teegarden}, {Cline}, {Briggs}, {Paciesas}, {Pendleton}, {Fishman},
		{Kouveliotou}, {Meegan}, {Wilson}, \& {Lestrade}}]{Band1993}
	{Band} D. {et~al.}, 1993, \apj, 413, 281
	
	\bibitem[{{Barkov} {et~al}\mbox{.}(2018){Barkov}, {Kathirgamaraju}, {Luo},
		{Lyutikov}, \& {Giannios}}]{Barkov2018}
	{Barkov} M.~V., {Kathirgamaraju} A., {Luo} Y., {Lyutikov} M., {Giannios} D.,
	2018, ArXiv e-prints
	
	\bibitem[{{Beniamini}, {Dvorkin} \& {Silk}(2018){Beniamini}, {Dvorkin}, \&
		{Silk}}]{BDS2018}
	{Beniamini} P., {Dvorkin} I., {Silk} J., 2018, \mnras, 478, 1994
	
	\bibitem[{{Beniamini} {et~al}\mbox{.}(2018){Beniamini}, {Giannios}, {Younes},
		{van der Horst}, \& {Kouveliotou}}]{Beniamini2018}
	{Beniamini} P., {Giannios} D., {Younes} G., {van der Horst} A.~J.,
	{Kouveliotou} C., 2018, \mnras, 476, 5621
	
	\bibitem[{{Beniamini}, {Hotokezaka} \& {Piran}(2016){Beniamini}, {Hotokezaka},
		\& {Piran}}]{Beniamini2016a}
	{Beniamini} P., {Hotokezaka} K., {Piran} T., 2016, \apj, 832, 149
	
	\bibitem[{{Beniamini} \& {Nakar}(2019)}]{BN2018}
	{Beniamini} P., {Nakar} E., 2019, \mnras, 482, 5430
	
	\bibitem[{{Beniamini} {et~al}\mbox{.}(2015){Beniamini}, {Nava}, {Duran}, \&
		{Piran}}]{Beniamini2015}
	{Beniamini} P., {Nava} L., {Duran} R.~B., {Piran} T., 2015, \mnras, 454, 1073
	
	\bibitem[{{Beniamini}, {Nava} \& {Piran}(2016){Beniamini}, {Nava}, \&
		{Piran}}]{Beniamini2016}
	{Beniamini} P., {Nava} L., {Piran} T., 2016, \mnras, 461, 51
	
	\bibitem[{{Blandford} \& {Znajek}(1977)}]{BZ1977}
	{Blandford} R.~D., {Znajek} R.~L., 1977, \mnras, 179, 433
	
	\bibitem[{{Blinnikov} {et~al}\mbox{.}(1984){Blinnikov}, {Novikov},
		{Perevodchikova}, \& {Polnarev}}]{Blinnikov1984}
	{Blinnikov} S.~I., {Novikov} I.~D., {Perevodchikova} T.~V., {Polnarev} A.~G.,
	1984, Soviet Astronomy Letters, 10, 177
	
	\bibitem[{{Bromberg} {et~al}\mbox{.}(2011){Bromberg}, {Nakar}, {Piran}, \&
		{Sari}}]{Bromberg2011}
	{Bromberg} O., {Nakar} E., {Piran} T., {Sari} R., 2011, \apj, 740, 100
	
	\bibitem[{{Bromberg} {et~al}\mbox{.}(2018){Bromberg}, {Tchekhovskoy},
		{Gottlieb}, {Nakar}, \& {Piran}}]{Bromberg2018}
	{Bromberg} O., {Tchekhovskoy} A., {Gottlieb} O., {Nakar} E., {Piran} T., 2018,
	\mnras, 475, 2971
	
	\bibitem[{{Burns} {et~al}\mbox{.}(2016){Burns}, {Connaughton}, {Zhang}, {Lien},
		{Briggs}, {Goldstein}, {Pelassa}, \& {Troja}}]{Burns2016}
	{Burns} E., {Connaughton} V., {Zhang} B.-B., {Lien} A., {Briggs} M.~S.,
	{Goldstein} A., {Pelassa} V., {Troja} E., 2016, \apj, 818, 110
	
	\bibitem[{{Coulter} {et~al}\mbox{.}(2017){Coulter}, {Foley}, {Kilpatrick},
		{Drout}, {Piro}, {Shappee}, {Siebert}, {Simon}, {Ulloa}, {Kasen}, {Madore},
		{Murguia-Berthier}, {Pan}, {Prochaska}, {Ramirez-Ruiz}, {Rest}, \&
		{Rojas-Bravo}}]{Coulter2017}
	{Coulter} D.~A. {et~al.}, 2017, Science, 358, 1556
	
	\bibitem[{{Coward} {et~al}\mbox{.}(2012){Coward}, {Howell}, {Piran}, {Stratta},
		{Branchesi}, {Bromberg}, {Gendre}, {Burman}, \& {Guetta}}]{Coward2012}
	{Coward} D.~M. {et~al.}, 2012, \mnras, 425, 2668
	
	\bibitem[{{Duffell} {et~al}\mbox{.}(2018){Duffell}, {Quataert}, {Kasen}, \&
		{Klion}}]{Duffell2018}
	{Duffell} P.~C., {Quataert} E., {Kasen} D., {Klion} H., 2018, \apj, 866, 3
	
	\bibitem[{{Eichler} \& {Levinson}(2004)}]{Eichler2004}
	{Eichler} D., {Levinson} A., 2004, \apjl, 614, L13
	
	\bibitem[{{Eichler} {et~al}\mbox{.}(1989){Eichler}, {Livio}, {Piran}, \&
		{Schramm}}]{Eichler1989}
	{Eichler} D., {Livio} M., {Piran} T., {Schramm} D.~N., 1989, \nat, 340, 126
	
	\bibitem[{{Finstad} {et~al}\mbox{.}(2018){Finstad}, {De}, {Brown}, {Berger}, \&
		{Biwer}}]{Finstad2018}
	{Finstad} D., {De} S., {Brown} D.~A., {Berger} E., {Biwer} C.~M., 2018, \apjl,
	860, L2
	
	\bibitem[{{Fong} {et~al}\mbox{.}(2015){Fong}, {Berger}, {Margutti}, \&
		{Zauderer}}]{Fong2015}
	{Fong} W., {Berger} E., {Margutti} R., {Zauderer} B.~A., 2015, \apj, 815, 102
	
	\bibitem[{{Fraija} {et~al}\mbox{.}(2017){Fraija}, {De Colle}, {Veres},
		{Dichiara}, {Barniol Duran}, \& {Galvan-Gamez}}]{Fraija2017}
	{Fraija} N., {De Colle} F., {Veres} P., {Dichiara} S., {Barniol Duran} R.,
	{Galvan-Gamez} A., 2017, ArXiv e-prints
	
	\bibitem[{{Frail} {et~al}\mbox{.}(2001){Frail}, {Kulkarni}, {Sari},
		{Djorgovski}, {Bloom}, {Galama}, {Reichart}, {Berger}, {Harrison}, {Price},
		{Yost}, {Diercks}, {Goodrich}, \& {Chaffee}}]{Frail2001}
	{Frail} D.~A. {et~al.}, 2001, \apjl, 562, L55
	
	\bibitem[{{Gehrels}(1986)}]{Gehrels1986}
	{Gehrels} N., 1986, \apj, 303, 336
	
	\bibitem[{{Gehrels}(2004)}]{Gehrels2004}
	{Gehrels} N., 2004, in American Institute of Physics Conference Series, Vol.
	727, Gamma-Ray Bursts: 30 Years of Discovery, {Fenimore} E., {Galassi} M.,
	eds., pp. 637--641
	
	\bibitem[{{Ghirlanda} {et~al}\mbox{.}(2016){Ghirlanda}, {Salafia}, {Pescalli},
		{Ghisellini}, {Salvaterra}, {Chassande-Mottin}, {Colpi}, {Nappo}, {D'Avanzo},
		{Melandri}, {Bernardini}, {Branchesi}, {Campana}, {Ciolfi}, {Covino},
		{G{\"o}tz}, {Vergani}, {Zennaro}, \& {Tagliaferri}}]{Ghirlanda2016}
	{Ghirlanda} G. {et~al.}, 2016, \aap, 594, A84
	
	\bibitem[{{Giannios}(2007)}]{Giannios2007}
	{Giannios} D., 2007, ArXiv e-prints
	
	\bibitem[{{Goldstein} {et~al}\mbox{.}(2017){Goldstein}, {Veres}, {Burns},
		{Briggs}, {Hamburg}, {Kocevski}, {Wilson-Hodge}, {Preece}, {Poolakkil},
		{Roberts}, {Hui}, {Connaughton}, {Racusin}, {von Kienlin}, {Dal Canton},
		{Christensen}, {Littenberg}, {Siellez}, {Blackburn}, {Broida}, {Bissaldi},
		{Cleveland}, {Gibby}, {Giles}, {Kippen}, {McBreen}, {McEnery}, {Meegan},
		{Paciesas}, \& {Stanbro}}]{Goldstein2017}
	{Goldstein} A. {et~al.}, 2017, \apjl, 848, L14
	
	\bibitem[{{Goodman}(1986)}]{Goodman1986}
	{Goodman} J., 1986, \apjl, 308, L47
	
	\bibitem[{{Gottlieb} {et~al}\mbox{.}(2017){Gottlieb}, {Nakar}, {Piran}, \&
		{Hotokezaka}}]{Gottlieb2017}
	{Gottlieb} O., {Nakar} E., {Piran} T., {Hotokezaka} K., 2017, ArXiv e-prints
	
	\bibitem[{{Granot} {et~al}\mbox{.}(2017){Granot}, {Gill}, {Guetta}, \& {De
			Colle}}]{Granot2017}
	{Granot} J., {Gill} R., {Guetta} D., {De Colle} F., 2017, ArXiv e-prints
	
	\bibitem[{{Granot}, {Guetta} \& {Gill}(2017){Granot}, {Guetta}, \&
		{Gill}}]{GGG2017}
	{Granot} J., {Guetta} D., {Gill} R., 2017, \apjl, 850, L24
	
	\bibitem[{{Granot} \& {Kumar}(2003)}]{GK2003}
	{Granot} J., {Kumar} P., 2003, \apj, 591, 1086
	
	\bibitem[{{Granot} \& {Ramirez-Ruiz}(2010)}]{GranotRamirez2010}
	{Granot} J., {Ramirez-Ruiz} E., 2010, ArXiv e-prints
	
	\bibitem[{{Guetta} \& {Piran}(2006)}]{Guetta2006}
	{Guetta} D., {Piran} T., 2006, \aap, 453, 823
	
	\bibitem[{{Guetta} \& {Stella}(2009)}]{Guetta2009}
	{Guetta} D., {Stella} L., 2009, \aap, 498, 329
	
	\bibitem[{{Hilborn}(2018)}]{Hilborn2018}
	{Hilborn} R.~C., 2018, ArXiv e-prints
	
	\bibitem[{{Hotokezaka}, {Beniamini} \& {Piran}(2018){Hotokezaka}, {Beniamini},
		\& {Piran}}]{HBP2018}
	{Hotokezaka} K., {Beniamini} P., {Piran} T., 2018, International Journal of
	Modern Physics D, 27, 1842005
	
	\bibitem[{{Hotokezaka} {et~al}\mbox{.}(2013){Hotokezaka}, {Kiuchi}, {Kyutoku},
		{Okawa}, {Sekiguchi}, {Shibata}, \& {Taniguchi}}]{Hotokezaka2013}
	{Hotokezaka} K., {Kiuchi} K., {Kyutoku} K., {Okawa} H., {Sekiguchi} Y.-i.,
	{Shibata} M., {Taniguchi} K., 2013, \prd, 87, 024001
	
	\bibitem[{{Hotokezaka} {et~al}\mbox{.}(2018){Hotokezaka}, {Kiuchi}, {Shibata},
		{Nakar}, \& {Piran}}]{Hotokezaka2018}
	{Hotokezaka} K., {Kiuchi} K., {Shibata} M., {Nakar} E., {Piran} T., 2018, ArXiv
	e-prints
	
	\bibitem[{{Hotokezaka}, {Piran} \& {Paul}(2015){Hotokezaka}, {Piran}, \&
		{Paul}}]{Hotokezaka2015}
	{Hotokezaka} K., {Piran} T., {Paul} M., 2015, Nature Physics, 11, 1042
	
	\bibitem[{{Ioka} \& {Nakamura}(2018)}]{Ioka2018}
	{Ioka} K., {Nakamura} T., 2018, Progress of Theoretical and Experimental
	Physics, 2018, 043E02
	
	\bibitem[{{Janka} {et~al}\mbox{.}(2006){Janka}, {Aloy}, {Mazzali}, \&
		{Pian}}]{Janka2006}
	{Janka} H.-T., {Aloy} M.-A., {Mazzali} P.~A., {Pian} E., 2006, \apj, 645, 1305
	
	\bibitem[{{Ji} {et~al}\mbox{.}(2016){Ji}, {Frebel}, {Chiti}, \&
		{Simon}}]{ji2016Nature}
	{Ji} A.~P., {Frebel} A., {Chiti} A., {Simon} J.~D., 2016, \nat, 531, 610
	
	\bibitem[{{Kasen} {et~al}\mbox{.}(2017){Kasen}, {Metzger}, {Barnes},
		{Quataert}, \& {Ramirez-Ruiz}}]{Kasen2017}
	{Kasen} D., {Metzger} B., {Barnes} J., {Quataert} E., {Ramirez-Ruiz} E., 2017,
	\nat, 551, 80
	
	\bibitem[{{Kasliwal} {et~al}\mbox{.}(2017){Kasliwal}, {Nakar}, {Singer},
		{Kaplan}, {Cook}, {Van Sistine}, {Lau}, {Fremling}, {Gottlieb}, {Jencson},
		{Adams}, {Feindt}, {Hotokezaka}, {Ghosh}, {Perley}, {Yu}, {Piran}, {Allison},
		{Anupama}, {Balasubramanian}, {Bannister}, {Bally}, {Barnes}, {Barway},
		{Bellm}, {Bhalerao}, {Bhattacharya}, {Blagorodnova}, {Bloom}, {Brady},
		{Cannella}, {Chatterjee}, {Cenko}, {Cobb}, {Copperwheat}, {Corsi}, {De},
		{Dobie}, {Emery}, {Evans}, {Fox}, {Frail}, {Frohmaier}, {Goobar}, {Hallinan},
		{Harrison}, {Helou}, {Hinderer}, {Ho}, {Horesh}, {Ip}, {Itoh}, {Kasen},
		{Kim}, {Kuin}, {Kupfer}, {Lynch}, {Madsen}, {Mazzali}, {Miller}, {Mooley},
		{Murphy}, {Ngeow}, {Nichols}, {Nissanke}, {Nugent}, {Ofek}, {Qi}, {Quimby},
		{Rosswog}, {Rusu}, {Sadler}, {Schmidt}, {Sollerman}, {Steele}, {Williamson},
		{Xu}, {Yan}, {Yatsu}, {Zhang}, \& {Zhao}}]{Kasliwal2017}
	{Kasliwal} M.~M. {et~al.}, 2017, Science, 358, 1559
	
	\bibitem[{{Kathirgamaraju}, {Barniol Duran} \&
		{Giannios}(2018){Kathirgamaraju}, {Barniol Duran}, \&
		{Giannios}}]{Kathirgamaraju2018}
	{Kathirgamaraju} A., {Barniol Duran} R., {Giannios} D., 2018, \mnras, 473, L121
	
	\bibitem[{{Kawanaka}, {Piran} \& {Krolik}(2013){Kawanaka}, {Piran}, \&
		{Krolik}}]{Kawanaka2013}
	{Kawanaka} N., {Piran} T., {Krolik} J.~H., 2013, \apj, 766, 31
	
	\bibitem[{{Kim}, {Perera} \& {McLaughlin}(2015){Kim}, {Perera}, \&
		{McLaughlin}}]{Kim2015}
	{Kim} C., {Perera} B.~B.~P., {McLaughlin} M.~A., 2015, \mnras, 448, 928
	
	\bibitem[{{Kochanek} \& {Piran}(1993)}]{Kochanek1993}
	{Kochanek} C.~S., {Piran} T., 1993, \apjl, 417, L17
	
	\bibitem[{{Kouveliotou} {et~al}\mbox{.}(1993){Kouveliotou}, {Meegan},
		{Fishman}, {Bhat}, {Briggs}, {Koshut}, {Paciesas}, \&
		{Pendleton}}]{Kouveliotou1993}
	{Kouveliotou} C., {Meegan} C.~A., {Fishman} G.~J., {Bhat} N.~P., {Briggs}
	M.~S., {Koshut} T.~M., {Paciesas} W.~S., {Pendleton} G.~N., 1993, \apjl, 413,
	L101
	
	\bibitem[{{Lamb} \& {Kobayashi}(2017)}]{Lamb2017}
	{Lamb} G.~P., {Kobayashi} S., 2017, \mnras, 472, 4953
	
	\bibitem[{{Lattimer} \& {Schramm}(1974)}]{Lattimer1974}
	{Lattimer} J.~M., {Schramm} D.~N., 1974, \apjl, 192, L145
	
	\bibitem[{{Lattimer} \& {Schramm}(1976)}]{Lattimer1976}
	{Lattimer} J.~M., {Schramm} D.~N., 1976, \apj, 210, 549
	
	\bibitem[{{Lazzati} {et~al}\mbox{.}(2017{\natexlab{a}}){Lazzati}, {Deich},
		{Morsony}, \& {Workman}}]{Lazzati2017B}
	{Lazzati} D., {Deich} A., {Morsony} B.~J., {Workman} J.~C., 2017{\natexlab{a}},
	\mnras, 471, 1652
	
	\bibitem[{{Lazzati} {et~al}\mbox{.}(2017{\natexlab{b}}){Lazzati},
		{L{\'o}pez-C{\'a}mara}, {Cantiello}, {Morsony}, {Perna}, \&
		{Workman}}]{Lazzati2017}
	{Lazzati} D., {L{\'o}pez-C{\'a}mara} D., {Cantiello} M., {Morsony} B.~J.,
	{Perna} R., {Workman} J.~C., 2017{\natexlab{b}}, \apjl, 848, L6
	
	\bibitem[{{Lazzati} {et~al}\mbox{.}(2018){Lazzati}, {Perna}, {Morsony},
		{Lopez-Camara}, {Cantiello}, {Ciolfi}, {Giacomazzo}, \&
		{Workman}}]{Lazzati2018}
	{Lazzati} D., {Perna} R., {Morsony} B.~J., {Lopez-Camara} D., {Cantiello} M.,
	{Ciolfi} R., {Giacomazzo} B., {Workman} J.~C., 2018, Physical Review Letters,
	120, 241103
	
	\bibitem[{{Lipunov}, {Postnov} \& {Prokhorov}(2001){Lipunov}, {Postnov}, \&
		{Prokhorov}}]{Lipunov2001}
	{Lipunov} V.~M., {Postnov} K.~A., {Prokhorov} M.~E., 2001, Astronomy Reports,
	45, 236
	
	\bibitem[{{Lu} {et~al}\mbox{.}(2017){Lu}, {Du}, {Cheng}, {L{\"u}}, {Zhang},
		{Lan}, \& {Liang}}]{Lu2017}
	{Lu} R.-J., {Du} S.-S., {Cheng} J.-G., {L{\"u}} H.-J., {Zhang} H.-M., {Lan} L.,
	{Liang} E.-W., 2017, ArXiv e-prints
	
	\bibitem[{{Macias} \& {Ramirez-Ruiz}(2016)}]{Macias2016}
	{Macias} P., {Ramirez-Ruiz} E., 2016, ArXiv e-prints
	
	\bibitem[{{Margutti} {et~al}\mbox{.}(2018){Margutti}, {Alexander}, {Xie},
		{Sironi}, {Metzger}, {Kathirgamaraju}, {Fong}, {Blanchard}, {Berger},
		{MacFadyen}, {Giannios}, {Guidorzi}, {Hajela}, {Chornock}, {Cowperthwaite},
		{Eftekhari}, {Nicholl}, {Villar}, {Williams}, \& {Zrake}}]{Margutti2018}
	{Margutti} R. {et~al.}, 2018, \apjl, 856, L18
	
	\bibitem[{{Moharana} \& {Piran}(2017)}]{Moharana2017}
	{Moharana} R., {Piran} T., 2017, \mnras, 472, L55
	
	\bibitem[{{Mooley} {et~al}\mbox{.}(2018{\natexlab{a}}){Mooley}, {Deller},
		{Gottlieb}, {Nakar}, {Hallinan}, {Bourke}, {Frail}, {Horesh}, {Corsi}, \&
		{Hotokezaka}}]{Mooley2018B}
	{Mooley} K.~P. {et~al.}, 2018{\natexlab{a}}, \nat, 561, 355
	
	\bibitem[{{Mooley} {et~al}\mbox{.}(2018{\natexlab{b}}){Mooley}, {Nakar},
		{Hotokezaka}, {Hallinan}, {Corsi}, {Frail}, {Horesh}, {Murphy}, {Lenc},
		{Kaplan}, {de}, {Dobie}, {Chandra}, {Deller}, {Gottlieb}, {Kasliwal},
		{Kulkarni}, {Myers}, {Nissanke}, {Piran}, {Lynch}, {Bhalerao}, {Bourke},
		{Bannister}, \& {Singer}}]{Mooley2018}
	{Mooley} K.~P. {et~al.}, 2018{\natexlab{b}}, \nat, 554, 207
	
	\bibitem[{{Murguia-Berthier} {et~al}\mbox{.}(2014){Murguia-Berthier}, {Montes},
		{Ramirez-Ruiz}, {De Colle}, \& {Lee}}]{Murguia-Berthier2014}
	{Murguia-Berthier} A., {Montes} G., {Ramirez-Ruiz} E., {De Colle} F., {Lee}
	W.~H., 2014, \apjl, 788, L8
	
	\bibitem[{{Nagakura} {et~al}\mbox{.}(2014){Nagakura}, {Hotokezaka},
		{Sekiguchi}, {Shibata}, \& {Ioka}}]{Nagakura2014}
	{Nagakura} H., {Hotokezaka} K., {Sekiguchi} Y., {Shibata} M., {Ioka} K., 2014,
	\apjl, 784, L28
	
	\bibitem[{{Nakar} \& {Piran}(2017)}]{NP2017}
	{Nakar} E., {Piran} T., 2017, \apj, 834, 28
	
	\bibitem[{{Nakar} \& {Sari}(2010)}]{SN2010}
	{Nakar} E., {Sari} R., 2010, \apj, 725, 904
	
	\bibitem[{{Nava} {et~al}\mbox{.}(2011){Nava}, {Ghirlanda}, {Ghisellini}, \&
		{Celotti}}]{Nava2011}
	{Nava} L., {Ghirlanda} G., {Ghisellini} G., {Celotti} A., 2011, \aap, 530, A21
	
	\bibitem[{{Paczynski}(1986)}]{Paczynski1986}
	{Paczynski} B., 1986, \apjl, 308, L43
	
	\bibitem[{{Patricelli} {et~al}\mbox{.}(2016){Patricelli}, {Razzano}, {Cella},
		{Fidecaro}, {Pian}, {Branchesi}, \& {Stamerra}}]{Patricelli2016}
	{Patricelli} B., {Razzano} M., {Cella} G., {Fidecaro} F., {Pian} E.,
	{Branchesi} M., {Stamerra} A., 2016, JCAP, 11, 056
	
	\bibitem[{{Pescalli} {et~al}\mbox{.}(2015){Pescalli}, {Ghirlanda}, {Salafia},
		{Ghisellini}, {Nappo}, \& {Salvaterra}}]{Pescalli2015}
	{Pescalli} A., {Ghirlanda} G., {Salafia} O.~S., {Ghisellini} G., {Nappo} F.,
	{Salvaterra} R., 2015, \mnras, 447, 1911
	
	\bibitem[{{Petropoulou}, {Barniol Duran} \& {Giannios}(2017){Petropoulou},
		{Barniol Duran}, \& {Giannios}}]{mariaetal2017}
	{Petropoulou} M., {Barniol Duran} R., {Giannios} D., 2017, \mnras, 472, 2722
	
	\bibitem[{{Pooley} {et~al}\mbox{.}(2018){Pooley}, {Kumar}, {Wheeler}, \&
		{Grossan}}]{Pooley2018}
	{Pooley} D., {Kumar} P., {Wheeler} J.~C., {Grossan} B., 2018, \apjl, 859, L23
	
	\bibitem[{{Racusin} {et~al}\mbox{.}(2017){Racusin}, {Burns}, {Goldstein},
		{Connaughton}, {Wilson-Hodge}, {Jenke}, {Blackburn}, {Briggs}, {Broida},
		{Camp}, {Christensen}, {Hui}, {Littenberg}, {Shawhan}, {Singer}, {Veitch},
		{Bhat}, {Cleveland}, {Fitzpatrick}, {Gibby}, {von Kienlin}, {McBreen},
		{Mailyan}, {Meegan}, {Paciesas}, {Preece}, {Roberts}, {Stanbro}, {Veres},
		{Zhang}, {Fermi LAT Collaboration}, {Ackermann}, {Albert}, {Atwood},
		{Axelsson}, {Baldini}, {Ballet}, {Barbiellini}, {Baring}, {Bastieri},
		{Bellazzini}, {Bissaldi}, {Blandford}, {Bloom}, {Bonino}, {Bregeon}, {Bruel},
		{Buson}, {Caliandro}, {Cameron}, {Caputo}, {Caragiulo}, {Caraveo},
		{Cavazzuti}, {Charles}, {Chiang}, {Ciprini}, {Costanza}, {Cuoco}, {Cutini},
		{D'Ammando}, {de Palma}, {Desiante}, {Digel}, {Di Lalla}, {Di Mauro}, {Di
			Venere}, {Drell}, {Favuzzi}, {Ferrara}, {Focke}, {Fukazawa}, {Funk}, {Fusco},
		{Gargano}, {Gasparrini}, {Giglietto}, {Gill}, {Giroletti}, {Glanzman},
		{Granot}, {Green}, {Grove}, {Guillemot}, {Guiriec}, {Harding}, {Jogler},
		{J{\'o}hannesson}, {Kamae}, {Kensei}, {Kocevski}, {Kuss}, {Larsson},
		{Latronico}, {Li}, {Longo}, {Loparco}, {Lubrano}, {Magill}, {Maldera},
		{Malyshev}, {Mazziotta}, {McEnery}, {Michelson}, {Mizuno}, {Monzani},
		{Morselli}, {Moskalenko}, {Negro}, {Nuss}, {Omodei}, {Orienti}, {Orlando},
		{Ormes}, {Paneque}, {Perkins}, {Pesce-Rollins}, {Piron}, {Pivato}, {Porter},
		{Principe}, {Rain{\`o}}, {Rando}, {Razzano}, {Razzaque}, {Reimer}, {Reimer},
		{Saz Parkinson}, {Scargle}, {Sgr{\`o}}, {Simone}, {Siskind}, {Smith},
		{Spada}, {Spinelli}, {Suson}, {Tajima}, {Thayer}, {Torres}, {Troja},
		{Uchiyama}, {Vianello}, {Wood}, \& {Wood}}]{Racusin2017}
	{Racusin} J.~L. {et~al.}, 2017, \apj, 835, 82
	
	\bibitem[{{Rossi}, {Lazzati} \& {Rees}(2002){Rossi}, {Lazzati}, \&
		{Rees}}]{Rossi2002}
	{Rossi} E., {Lazzati} D., {Rees} M.~J., 2002, \mnras, 332, 945
	
	\bibitem[{{Salvaterra} {et~al}\mbox{.}(2008){Salvaterra}, {Cerutti},
		{Chincarini}, {Colpi}, {Guidorzi}, \& {Romano}}]{Salvaterra2008}
	{Salvaterra} R., {Cerutti} A., {Chincarini} G., {Colpi} M., {Guidorzi} C.,
	{Romano} P., 2008, \mnras, 388, L6
	
	\bibitem[{{Schutz}(2011)}]{Schutz2011}
	{Schutz} B.~F., 2011, Classical and Quantum Gravity, 28, 125023
	
	\bibitem[{{Sobacchi} {et~al}\mbox{.}(2017){Sobacchi}, {Granot}, {Bromberg}, \&
		{Sormani}}]{Sobacchi2017}
	{Sobacchi} E., {Granot} J., {Bromberg} O., {Sormani} M.~C., 2017, \mnras, 472,
	616
	
	\bibitem[{{Tanaka} {et~al}\mbox{.}(2017){Tanaka}, {Utsumi}, {Mazzali},
		{Tominaga}, {Yoshida}, {Sekiguchi}, {Morokuma}, {Motohara}, {Ohta},
		{Kawabata}, {Abe}, {Aoki}, {Asakura}, {Baar}, {Barway}, {Bond}, {Doi},
		{Fujiyoshi}, {Furusawa}, {Honda}, {Itoh}, {Kawabata}, {Kawai}, {Kim}, {Lee},
		{Miyazaki}, {Morihana}, {Nagashima}, {Nagayama}, {Nakaoka}, {Nakata},
		{Ohsawa}, {Ohshima}, {Okita}, {Saito}, {Sumi}, {Tajitsu}, {Takahashi},
		{Takayama}, {Tamura}, {Tanaka}, {Terai}, {Tristram}, {Yasuda}, \&
		{Zenko}}]{Tanaka2017}
	{Tanaka} M. {et~al.}, 2017, \pasj, 69, 102
	
	\bibitem[{{Tanvir} {et~al}\mbox{.}(2017){Tanvir}, {Levan},
		{Gonz{\'a}lez-Fern{\'a}ndez}, {Korobkin}, {Mandel}, {Rosswog}, {Hjorth},
		{D'Avanzo}, {Fruchter}, {Fryer}, {Kangas}, {Milvang-Jensen}, {Rosetti},
		{Steeghs}, {Wollaeger}, {Cano}, {Copperwheat}, {Covino}, {D'Elia}, {de Ugarte
			Postigo}, {Evans}, {Even}, {Fairhurst}, {Figuera Jaimes}, {Fontes}, {Fujii},
		{Fynbo}, {Gompertz}, {Greiner}, {Hodosan}, {Irwin}, {Jakobsson},
		{J{\o}rgensen}, {Kann}, {Lyman}, {Malesani}, {McMahon}, {Melandri},
		{O'Brien}, {Osborne}, {Palazzi}, {Perley}, {Pian}, {Piranomonte}, {Rabus},
		{Rol}, {Rowlinson}, {Schulze}, {Sutton}, {Th{\"o}ne}, {Ulaczyk}, {Watson},
		{Wiersema}, \& {Wijers}}]{Tanvir2017}
	{Tanvir} N.~R. {et~al.}, 2017, \apjl, 848, L27
	
	\bibitem[{{Troja} {et~al}\mbox{.}(2018){Troja}, {Piro}, {Ryan}, {van Eerten},
		{Ricci}, {Wieringa}, {Lotti}, {Sakamoto}, \& {Cenko}}]{Troja2018}
	{Troja} E. {et~al.}, 2018, \mnras, 478, L18
	
	\bibitem[{{van Eerten} {et~al}\mbox{.}(2018){van Eerten}, {Ryan}, {Ricci},
		{Burgess}, {Wieringa}, {Piro}, {Cenko}, \& {Sakamoto}}]{vanEerten2018}
	{van Eerten} E.~T.~H., {Ryan} G., {Ricci} R., {Burgess} J.~M., {Wieringa} M.,
	{Piro} L., {Cenko} S.~B., {Sakamoto} T., 2018, ArXiv e-prints
	
	\bibitem[{{van Eerten} \& {MacFadyen}(2012)}]{VanEerten2012}
	{van Eerten} H.~J., {MacFadyen} A.~I., 2012, \apj, 751, 155
	
	\bibitem[{{Wanderman} \& {Piran}(2010)}]{WP2010}
	{Wanderman} D., {Piran} T., 2010, \mnras, 406, 1944
	
	\bibitem[{{Wanderman} \& {Piran}(2015)}]{wanderman_piran2015}
	{Wanderman} D., {Piran} T., 2015, \mnras, 448, 3026
	
	\bibitem[{{Zhang} \& {M{\'e}sz{\'a}ros}(2002)}]{Zhang2002}
	{Zhang} B., {M{\'e}sz{\'a}ros} P., 2002, \apj, 571, 876
	
	\bibitem[{{Zhang} {et~al}\mbox{.}(2018){Zhang}, {Zhang}, {Sun}, {Lei}, {Gao},
		{Li}, {Shao}, {Zhao}, {Hu}, {L{\"u}}, {Wu}, {Fan}, {Wang}, {Castro-Tirado},
		{Zhang}, {Yu}, {Cao}, \& {Liang}}]{Zhang2018}
	{Zhang} B.-B. {et~al.}, 2018, Nature Communications, 9, 447
	
\end{thebibliography}
\end{document}